\documentclass[12pt]{article}
\usepackage[utf8]{inputenc}
\usepackage[left=0.875in, right=0.875in]{geometry}
\usepackage{amsmath, amssymb, stmaryrd}
\usepackage{graphicx}
\usepackage{placeins}
\usepackage{float}
\usepackage{tabu, booktabs, multicol}
\usepackage{todonotes}
\usepackage{setspace}
\usepackage[
	backend=biber,
    style=ieee,
	sorting=none
]{biblatex}
\title{On the effective magnetostrictive properties of anisotropic magneto-active elastomers in the small-deformation limit}
\author{Connor D. Pierce\textsuperscript{a} and Kathryn H. Matlack\textsuperscript{a,*} \\
\small \textsuperscript{a}University of Illinois at Urbana-Champaign, Department of Mechanical Science and Engineering,\\
\small 1206 West Green Street, Urbana, IL, 61801, USA\\[0.125in]
\small \textsuperscript{*} Corresponding author\\[0.125in]
\small Email addresses: cpierce3@illinois.edu (C. Pierce), kmatlack@illinois.edu (K. Matlack)}
\date{}

\begin{document}

\maketitle

\begin{abstract}
Magneto-active elastomers (MAEs) are composite materials comprising an elastomer matrix with embedded magnetic particles, endowing the composite with coupled effective magneto-mechanical responses. MAEs can be prepared in isotropic formulations with random particle distributions or in anisotropic formulations with chain-like structures of particles, and it is widely reported that anisotropic MAEs exhibit much stronger magneto-mechanical coupling than isotropic MAEs. However, most efforts to model the effective magneto-mechanical properties of MAEs via homogenization approaches have focused on isotropic microstructures or microstructures which are constructed to have large separations between particles, in order to make use of analytical homogenization solutions. Such models therefore exclude by construction the microstructural effects which contribute to the enhanced magneto-mechanical coupling of chain-like anisotropic MAEs. In this work, we introduce a periodic homogenization approach to compute the effective magneto-mechanical properties of anisotropic MAEs, and use this model to analyze the microstructural features enhance their magneto-mechanical coupling. Using the finite element method, we numerically determine the effect of various microstructural parameters---including the particle shape, gap between particles, and presence of voids in the microstructure---on the effective stiffness, effective permeability, and effective magneto-mechanical coupling tensors for chain-like periodic microstructures in three dimensions. Using insights gained from the full-field simulations, we derive an analytical expression for the magneto-mechanical coupling in the chain direction, in terms of the volume fraction, gap size, and properties of the matrix and particles. Our results show that the overall magnetostriction of anisotropic MAEs is most sensitive to the gap between particles and the waviness of the particle chains, with smaller gap sizes and straighter chains yielding higher overall magnetostriction. Interestingly, our simulations show that while isotropic MAEs elongate in a uniform magnetic field, anisotropic MAEs contract with much larger strain amplitudes, a result of the attractive forces between particles being much stronger in anisotropic MAEs than in isotropic MAEs because of the close particle spacing found in anisotropic MAEs. Our results provide fundamental insights into the mechanisms that govern magneto-mechanical coupling in anisotropic MAEs, and additionally constitute a toolbox of material properties for design of magnetostrictive devices.
\end{abstract}


\section{Introduction}

Magneto-active elastomers (MAEs), also known as magneto-rheological elastomers (MREs), are soft materials comprising magnetic inclusions embedded in an elastomer matrix. Typical formulations include spherical ferromagnetic particles in a silicone or natural rubber matrix, with particles either randomly dispersed (creating an isotropic MAE) or aligned into chains by an external magnetic field applied during curing (creating an anisotropic MAE). The inclusion of magnetic particles endows the MAE with magneto-responsive mechanical properties including magnetostriction (magnetically-induced deformation, e.g. \cite{Galipeau2013}) and the magneto-rheological (MR) effect (magnetically-induced change in stiffness, e.g. \cite{Walter2017}). It has been proposed that the magneto-mechanical coupling of MAEs could enable a host of smart devices actuated by a magnetic field \cite{Li2014a}, including adaptive vibration absorbers \cite{Liao2012}, tunable metamaterials \cite{Yu2018, Pierce2020}, or actuators for soft robots \cite{Zhao2022, Moreno-Mateos2022}.

MAEs with a random particle dispersion (henceforth denoted ``isotropic MAEs'') are mechanically isotropic and either magnetically isotropic (if soft-magnetic particles are used, e.g. \cite{Bodelot2018}) or magnetically anisotropic (if hard-magnetic particles are used and magnetized after the MAE is cured \cite{Wang2020b}). MAEs with particles aligned in chains (henceforth ``anisotropic MAEs'') are both mechanically and magnetically anisotropic, with the anisotropy arising from the spatially anisotropic distribution of the magnetic particles rather than from anisotropy of the constituent materials \cite{Bustamante2010a}. It is widely reported that anisotropic MAEs exhibit stronger magneto-mechanical coupling than isotropic MAEs, due to the close spacing of particles in the chains \cite{Chen2007}.

Developing constitutive models for the effective magneto-mechanical behavior of MAEs has been the focus of considerable research over the past two-and-a-half decades. These models generally adopt one of three approaches: micromechanical modeling, phenomenological macroscopic free-energy functions, or homogenization methods. Micromechanical models typically consider single particles or aggregations of a few particles, and attempt to derive one or more effective properties of the MAE based on some simplified relations for the magnetic interactions between the particles, e.g. \cite{Jolly1996, Davis1999}. While having the advantage of being predictive, these models typically fail to provide a complete set of properties describing the magneto-mechanical response of the MAE, making it difficult to use these models to predict the fully-coupled macroscopic MAE response. Phenomenological models adopt a different approach by describing the large-strain magneto-mechanical deformation of MAE in terms of a nonlinear macroscopic free-energy function, which is usually expressed in terms of certain invariants possessing isotropic or transversely isotropic symmetry \cite{Dorfmann2004, Bustamante2010a, Danas2012}. These models have the advantage that they possess a limited number of material constants or functions which can be fit to experimental data. However, because they do not involve any microstructural analysis of the MAE, they are unable to \emph{predict} the MAE behavior from the properties of its constituents and cannot be used to study the microstructural effects that govern and enhance the magneto-mechanical coupling of the MAE.

Homogenization approaches present an interesting opportunity to produce a full set of magneto-mechanical properties like phenomenological approaches, but with the predictive power of micromechanical models. Furthermore, because they explicitly consider the microstructure of the MAE, they can be used to study how different aspects of the MAE microstructure contribute to and/or enhance the MAE's macroscopic magneto-mechanical coupling. Over the last decade, several homogenization models have been proposed for MAEs and a related class of materials known as dielectric elastomer composites (DECs). Due to the mathematical similarity between the static electro-mechanical and static magneto-mechanical problems, the results for DECs can be applied to MAEs by substituting the appropriate magnetic quantities for the electrical ones. In \cite{Tian2012}, Tian et al. introduced a homogenization framework in the context of small strains which describes the macroscopic electrostrictive behavior of DECs in terms of three effective material tensors, and reported on the effective properties of hierarchical laminates. This framework was subsequently utilized in \cite{Lefevre2014} to predict the effective properties of DECs comprising an isotropic distribution of spherical particles. An alternative framework for DECs proposed in \cite{Lopez-Pamies2014} allows to consider deformations and magnetic fields of arbitrary magnitudes, and was subsequently specialized to the small-deformation limit for isotropic \cite{Spinelli2015} and transversely isotropic \cite{Lefevre2015} distributions of particles. More recently, a fully nonlinear implicit homogenization model originally developed for DECs \cite{Lefevre2017b} was explicitly applied to the context of MAEs \cite{Lefevre2017}, and subsequently approximated by an explicit model \cite{Lefevre2020a}.

Despite the recent progress in homogenization models for DECs and MAEs, few works have explicitly considered chain-like microstructures of particles, which are widely reported to enhance the magneto-mechanical coupling of anisotropic MAEs relative to isotropic MAEs \cite{Chen2007, }. These microstructures are interesting because enormous changes in the MAE's effective properties can be induced in the chain direction due to contact between neighboring particles, even when the overall volume fraction of particles is dilute. This contrasts with isotropic MAEs, in which such large changes in the effective properties can only be realized at volume fractions approaching percolation. It is therefore of great interest to study how the details of the chain-like microstructure affect the obtained effective properties. Some early micromechanical models of anisotropic MAEs have explicitly considered chains of magnetic particles \cite{Jolly1996, Davis1999}, but limited their study to the magnetically-induced increment of shear modulus. A more recent homogenization model of transversely isotropic MAEs \cite{Lefevre2015} provides a full anisotropic constitutive model, but is explicitly constructed to have large spacings between particles and thus excludes by construction the very mechanisms which may contribute to enhanced coupling in chain-like MAEs. Idealized and more realistic chain-like microstructures under finite strains were considered in \cite{Kalina2016, Danas2017}, but the analysis was limited to only a few components of the effective stress, strain and magnetic field, and a systematic study of microstructural parameters was not reported. Consequently, the effect of microstructural parameters on the macroscopic magneto-mechanical coupling of chain-like anisotropic MAEs remains poorly understood.

The objective of this paper is to bridge that gap by developing a homogenization model of the coupled magneto-mechanical behavior of MAEs with chain-like particulate microstructures, and to employ the homogenization model to systematically study the microstructural parameters which govern the effective magneto-mechanical properties. To this end, we employ the small-strain periodic homogenization framework of \cite{Tian2012} and numerically study the effective properties of anisotropic MAEs using the finite element method for unit cells which produce chain-like microstructures when tessellated in space. We systematically vary microstructural parameters such as volume fraction, particle-to-particle spacing, particle shape, and chain waviness in order to determine their effect on the material properties. Guided by insights gleaned from our simulations, we develop an analytical expression for the effective magneto-mechanical coupling coefficient in the direction of the particle chains. Finally, we consider the magnetic field-induced macroscopic strain in a simple geometry, and show how the interplay of various microstructural parameters affects the macroscopic magneto-mechanical response of anisotropic MAEs.

\section{Approach} \label{sec:thry}

\subsection{Homogenization model}
The problem at hand is to determine the overall effective properties of a magneto-responsive composite material with a microstructure comprising chains of magnetic inclusions. When strains are small and magnetic fields ``moderate'', this problem can be solved by a suitable application of the the periodic homogenization theory of Tian et al. \cite{Tian2012}. We begin with a brief review of the principal results of \cite{Tian2012}, adapting the notation to the present case of magneto-responsive solids.

Consider a deformable body occupying the volume \(\Omega\), comprising a distribution of ferromagnetic particles firmly bonded to an elastomeric matrix, and subject to an externally applied magnetic field as well as displacement and traction boundary conditions. We denote the volume occupied by the matrix as \(\Omega_{(1)}\) and the volume occupied by the particles as \(\Omega_{(2)}\). The distribution of particles may be random and isotropic (if no magnetic field was applied during curing of the elastomer matrix), or the particles may be arranged into chain-like structures that are roughly parallel to the direction of the magnetic field applied during curing, depending on the preparation of the material. Under the assumption of small deformations and moderate magnetic fields, i.e. that strains \(\boldsymbol{\varepsilon}\) are \(O(\zeta)\) and magnetic fields \(\boldsymbol{H}\) are \(O(\zeta^{1/2})\), where \(\zeta\) is a small parameter, and that the free energy is an even function with respect to the magnetic field (i.e. the constituent materials are magnetically soft), both the matrix and particles are governed by free-energy density functions of the form
\begin{align}
	W(\boldsymbol{\varepsilon}, \boldsymbol{H}) =&
		\frac{1}{2} L_{ijkl} \varepsilon_{ij} \varepsilon_{kl}
		+ \mu_0 M_{ijkl} \varepsilon_{ij} H_k H_l
		- \frac{1}{2} \mu_0 \mu_{ij} H_i H_j - \frac{1}{24}A_{ijkl} H_i H_j H_k H_l + O\bigl(\zeta^{3/2}\bigr)\,. \label{eq:thry-free-energy}
\end{align}
In (\ref{eq:thry-free-energy}), \(\boldsymbol{L}\) is the familiar fourth-order stiffness tensor with major and minor symmetries, \(\mu_0\) is the permeability of free space, \(\boldsymbol{\mu}\) is the symmetric second-order relative permeability tensor, and \(\boldsymbol{M}\) is a fourth-order tensor that governs the magneto-mechanical coupling of the body. The fourth-order tensor \(\boldsymbol{A}\) determines the higher-order dependence of the free energy on the magnetic field, but as we shall see shortly, it does not factor into the constitutive relation. While \(\boldsymbol{L}\) possesses both major and minor symmetries, the existence of a free-energy function requires only that \(\boldsymbol{M}\) possess minor symmetries, i.e. \(M_{ijkl} = M_{jikl} = M_{ijlk}\). We note that while some authors embed the constant \(\mu_0\) in the tensors \(\boldsymbol{\mu}\) and \(\boldsymbol{M}\), we choose to factor it out of the tensors, so that \(\boldsymbol{\mu}\) and \(\boldsymbol{M}\) are dimensionless, while \(\boldsymbol{L}\) has dimensions of stress. From this free energy, the mechanical and magnetic constitutive relations of the body may be derived (to leading order):
\begin{subequations} \label{eq:thry-const-eq}
	\begin{align}
		\sigma_{ij} =& \frac{\mathrm{d}W}{\mathrm{d} \varepsilon_{ij}}
			 = L_{ijkl} \varepsilon_{kl} + \mu_0 M_{ijkl} H_k H_l \, , \\
		B_i =& -\frac{\mathrm{d}W}{\mathrm{d} H_i} = \mu_0 \mu_{ij} H_j \, ,
	\end{align}
\end{subequations}
where \(\boldsymbol{\sigma}\) and \(\boldsymbol{B}\) are the ``total'' stress tensor and the magnetic flux density, respectively, and no distinction is drawn between Cauchy and Piola-Kirchoff stresses because deformations are small. Equation~(\ref{eq:thry-const-eq}) shows that the problem is one-way coupled, i.e. that the magnetic fields induce stresses which can cause mechanical deformation, but the deformation does not influence the magnetic fields. The coupling is governed by the tensor \(\boldsymbol{M}\), which describes how the magnetic fields \(\boldsymbol{H}\) create a pre-stress which deforms the material. \(\boldsymbol{M}\) may be decomposed as \cite{Tian2012}
\begin{align}
	M_{ijkl} = M^{mat}_{ijkl} + \mu_{jl} \delta_{ik} - \frac{1}{2} \delta_{ij} \delta_{kl} \, , \label{eq:M-decomposition}
\end{align}
where \(\boldsymbol{M}^{mat}\) is the ``inherent'' material magnetostriction, and the remaining terms multiplied by \(\mu_0 H_k H_l\) give the Maxwell stress inside the material.

In \cite{Tian2012}, it was shown that the effective behavior of the composite is governed by the following partial differential equations (PDEs) for the macroscopic magnetic scalar potential \(\varphi\) and displacement \(\boldsymbol{u}\) fields, respectively:
\begin{subequations} \label{eq:thry-macro-gov}
	\begin{align}
		\nabla \cdot \left(\tilde{\boldsymbol{\mu}} \nabla \varphi \right) =& 0 & &\forall \boldsymbol{x}, \\
		\nabla \cdot \left(\tilde{\boldsymbol{L}} \nabla \boldsymbol{u} + \tilde{\boldsymbol{M}} \nabla \varphi \nabla \varphi \right) =& 0 & &\forall \boldsymbol{x} \in \Omega \, ,
	\end{align}
\end{subequations}
subject to appropriate boundary conditions. In (\ref{eq:thry-macro-gov}), \(\tilde{\boldsymbol{L}}\), \(\tilde{\boldsymbol{\mu}}\), and \(\tilde{\boldsymbol{M}}\) are the effective stiffness, permeability, and magneto-mechanical coupling tensors of the composite, respectively. It was further shown that for composites with periodic microstructures, the effective material tensors may be calculated from the following expressions:
\begin{subequations} \label{eq:thry-eff-tensors}
	\begin{align}
		\tilde{L}_{ijkl} =& \frac{1}{|Y|} \int_Y {
			L_{ijmn}(\boldsymbol{y}) G_{mnkl}(\boldsymbol{y}) \, \mathrm{d}Y
		} \, , \label{eq:thry-L-eff} \\
		\tilde{\mu}_{ij} =& \frac{1}{|Y|} \int_Y {
			\mu_{ik}(\boldsymbol{y}) g_{kj}(\boldsymbol{y}) \, \mathrm{d}Y
		} \, , \label{eq:thry-mu-eff} \\
		\tilde{M}_{ijkl} =& \frac{1}{|Y|} \int_Y {
			M_{mnpq}(\boldsymbol{y}) G_{mnij}(\boldsymbol{y}) g_{pk}(\boldsymbol{y}) g_{ql}(\boldsymbol{y}) \, \mathrm{d}Y
		} \, , \label{eq:thry-M-eff}
	\end{align}
\end{subequations}
where \(Y\) is the periodic unit cell that defines the microstructure and \(\boldsymbol{G}\) and \(\boldsymbol{g}\) are the (fourth-order) elastic concentration tensor and the (second-order) magnetic field concentration tensor, respectively. Hereafter, \(Y_{(1)}\) denotes the portion of the unit cell occupied by the matrix and \(Y_{(2)}\) denotes the portion occupied by the particle(s). The concentration tensors, defined as
\begin{align}
	G_{ijkl}(\boldsymbol{y}) =& \delta_{ik} \delta_{jl} + \frac{\partial \Gamma_{ikl}(\boldsymbol{y})}{\partial y_j} \, , &
	g_{ij}(\boldsymbol{y}) =& \delta_{ij} - \frac{\partial \gamma_j(\boldsymbol{y})}{\partial y_i} \, , \label{eq:thry-conc-tensors}
\end{align}
can be obtained from the solution of two uncoupled PDEs for \(\boldsymbol{\Gamma}\) and \(\boldsymbol{\gamma}\) on the unit cell, where \(\boldsymbol{\Gamma}\) and \(\boldsymbol{\gamma}\) are third- and first-order tensors with zero average on \(Y\), respectively, satisfying
\begin{subequations} \label{eq:thry-micro-pde}
	\begin{align}
		\frac{\partial}{\partial y_i} \left(
			\mu_{ik}(\boldsymbol{y}) \frac{\partial \gamma_j}{\partial y_k}
		\right) =& \frac{\partial \mu_{ij}(\boldsymbol{y})}{\partial y_i} & \forall \boldsymbol{y} \in Y \, , \label{eq:thry-micro-pde-g} \\
		- \frac{\partial}{\partial y_j} \left(
			L_{ijlm}(\boldsymbol{y}) \frac{\partial \Gamma_{lkh}}{\partial y_m}
		\right) =& \frac{\partial L_{ijkh}(\boldsymbol{y})}{\partial y_j} & \forall \boldsymbol{y} \in Y \, . \label{eq:thry-micro-pde-G}
	\end{align}
\end{subequations}
Note that \(\boldsymbol{\Gamma}\) is symmetric with respect to its second and third indices. To compute the effective tensors, one must identify a periodic unit cell \(Y\) defining the microstructure, compute the solution to the microscopic problem (\ref{eq:thry-micro-pde}), and compute the indicated volume averages (\ref{eq:thry-eff-tensors}). It is the objective of this paper to compute the effective properties tensors \(\boldsymbol{L}\), \(\boldsymbol{\mu}\), and \(\boldsymbol{M}\) for anisotropic MAEs, wherein the periodic unit cell \(Y\) defines an anisotropic geometry with parallel chains of closely-spaced particles, and to study how the geometric details of \(Y\) affect the effective properties.

\subsection{Finite element method implementation}
We implemented the homogenization problem (\ref{eq:thry-eff-tensors}-\ref{eq:thry-micro-pde}) computationally using the finite element method in COMSOL Multiphysics. We first created unit cell models capturing various geometric aspects of the microstructure as described in the following subsection, and meshed the model ensuring that the thin region between particles was spanned by at least two elements to ensure sufficient resolution of the solution, while also enforcing mesh periodicity on the unit cell boundary. We then solved the microscopic problem (\ref{eq:thry-micro-pde}) using a segregated approach, with the magnetic microscopic problem (\ref{eq:thry-micro-pde-g}) solved in one step using three scalar unknowns and the mechanical microscopic problem (\ref{eq:thry-micro-pde-G}) solved in a separate step using 18 scalar unknowns (taking into account the symmetry \(\Gamma_{ijk} = \Gamma_{ikj}\)). We used quadratic Lagrange-type shape functions in all simulations. In lieu of enforcing the zero-volume-average condition on \(\boldsymbol{\gamma}\) and \(\boldsymbol{\Gamma}\), we instead prescribed homogeneous Dirichlet boundary conditions on the solution at a single mesh vertex to make the problem well-posed. This avoided the need to use Lagrange multipliers to enforce the zero average, which would have introduced dense rows in the otherwise sparse system matrix. Since \(\boldsymbol{\gamma}\) and \(\boldsymbol{\Gamma}\) enter into the effective properties computation (\ref{eq:thry-eff-tensors}) only through their gradients, the constant offset introduced by substituting Dirichlet boundary conditions for the zero volume average condition does not affect the computed effective properties, which we verified by comparing with simulations in which the zero-volume-average condition was prescribed via Lagrange multipliers.

\subsection{Unit cell geometries}
\begin{figure}
	\centering
	\includegraphics[width=6.25in]{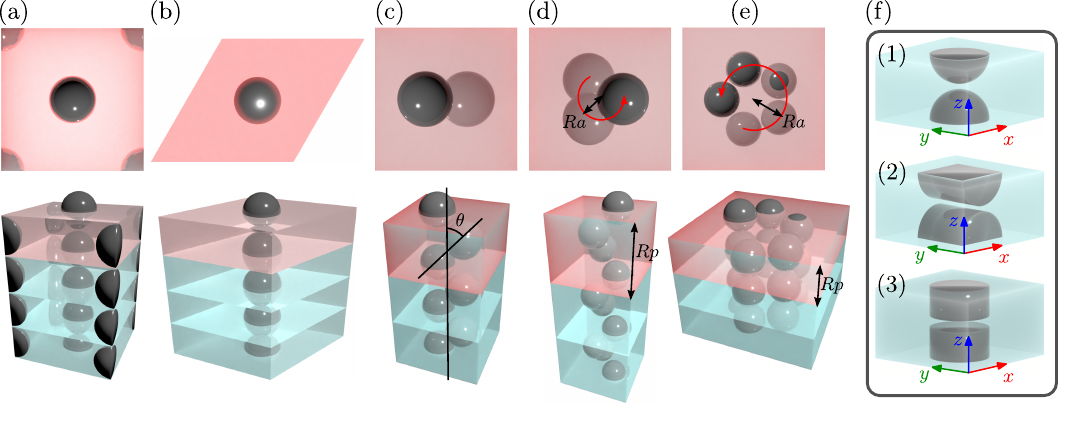}
	\caption{Unit cell geometries studied in this work: (a) TetC, (b) Hex, (c) TetPw with \(\theta=45^\circ\), (d) TetPs-3 with \(a=0.75a_{max}\), (e) TetPs-5 with \(a=a_{max}\), and (f) comparison of unit cells with different particle shapes: (f1) TetP, (f2) TetPc-2, and (f3) TetPc-3. For interpretation of the abbreviations used to describe the unit cells, the reader is referred to Table~\ref{tbl:methods-abbrev}.}
	\label{fig:unit-cells}
\end{figure}

To emulate the chain-like arrangement of particles in anisotropic MAE, we modeled unit cells having a thickness only slightly larger than the particle size, such that when the unit cell is repeated periodically, a chain of closely-spaced particles is obtained, as shown in Figure~\ref{fig:unit-cells}. In all unit cells studied, the chain direction is arbitrarily chosen to be parallel to the global \(z\)-axis. We define a parameter \(d\) which governs the particle-to-particle spacing, such that the gap between adjacent particles in the chain is \(2 l d\), where \(l\) is the characteristic length of the particle. (For spherical particles and cylindrical particles in the TetPc-2 orientation, \(l = R\), with \(R\) the particle radius. For TetPc-3 unit cells, \(l = L\), with \(L\) the cylinder height.) For straight chains of particles, the unit cell thickness is therefore fixed as \(2(1 + d)l\). To control the volume fraction \(c\), we varied the dimensions of the unit cell in the plane perpendicular to the chain (the \(xy\)-plane in all unit cells studied). The parameter \(d\) thus controls the size of the gap between adjacent particles in a chain, while the parameter \(c\) controls the spacing between chains. For each geometry, there exists a volume fraction \(c_{max}\) for which the chain-to-chain spacing in the \(xy\)-plane is equal to the particle-to-particle spacing in the \(z\)-direction. We call this the ``pseudo-percolation'' volume fraction for a given geometry, and we restrict our study to volume fractions \(c \leq c_{max}\).

Since anisotropic MAEs are obtained in practice by applying a magnetic field to an initially random distribution of particles suspended in liquid elastomer precursor, the distribution of particle chains in the \(xy\)-plane in a real MAE is random, endowing the MAE with transversely isotropic (TI) properties. Modeling anisotropic MAE microstructures with periodic unit cells thus introduces artificial symmetries that are not present in real anisotropic MAEs. To evaluate the effect of the unit cell symmetry on the effective properties, we studied unit cells with three different symmetries: tetragonal with primitive centering (``TetP''), tetragonal with body centering (``TetC'', Figure~\ref{fig:unit-cells}(a)), and hexagonal (``Hex'', Figure~\ref{fig:unit-cells}(b)). In addition, to evaluate how much the artificial symmetries affect the effective properties, and to enforce the expected TI symmetry on the obtained effective constitutive tensors, we performed a second homogenization step. In this step, we solved the homogenization problem (\ref{eq:thry-eff-tensors}-\ref{eq:thry-micro-pde}) for a geometry comprising a \(9\times 9\) grid of rectangular domains, where the effective properties for each domain were assigned as the effective properties from the first homogenization step with a randomized orientation in the \(xy\)-plane, illustrated schematically in Figure~\ref{fig:straight-chain-TI}(a-b). To evaluate how well the \(9\times 9\) grid approximated a true TI microstructure, we repeated this calculation for 20 unit cells with different randomized orientations.

Our homogenization approach enabled us to study a wide variety of physically-motivated microstructural parameters to evaluate their effects on the effective properties of anisotropic MAEs. In addition to the gap size, volume fraction, and unit cell symmetry, we also studied the particle chain waviness, particle shape, and presence of voids in the MAE. Since electron micrographs of anisotropic MAE microstructures generally reveal that particle chains are not straight as in Figure~\ref{fig:unit-cells}(a-b), but rather contain short-range waviness \cite{Chen2007,Boczkowska2012}, we studied unit cells representing wavy chains of spherical particles, as shown in Figure~\ref{fig:unit-cells}(c-e). The particle-to-particle spacing remained fixed at \(2 R d\), but the particles were arranged in zig-zag patterns at an angle \(\theta\) to the particle chain (``TetPw'', Figure~\ref{fig:unit-cells}(c)), or in spirals of radius \(R a\) and pitch \(R p\) with \(n\) particles per spiral revolution (``TetPs-\(n\)'', Figure~\ref{fig:unit-cells}(d-e)). While most experimental MAE studies utilize spherical particles, we studied the possible benefits of other particle shapes by modeling geometries with cylindrical and spheroidal particles. For cylindrical particles, we modeled unit cells with cylinders in different orientations, denoted ``TetPc-2'' and ``TetPc-3'' when the particle axis is parallel to the global \(y\)- and \(z\)-axes, respectively. For spheroidal particles, we aligned the spheroidal axis of revolution parallel to the chain axis, and assigned semi-radii of \(R\) in the chain direction and \(R\alpha\) in the orthogonal direction, where \(\alpha\) controls the particle aspect ratio. Thus, prolate spheroids are obtained for \(\alpha<1\) and oblate spheroids for \(\alpha>1\). These unit cells are denoted ``TetPe-\(\alpha\)''. Finally, we studied the effect of voids in particle chains (which could result from incomplete wetting of the particles by the matrix) by modeling a cylindrical inclusion \(Y_{(3)}\) in the gap between particles in a chain. In this inclusion, we prescribed mechanical and magnetic properties that correspond to vacuum, namely: very low stiffness \(\boldsymbol{L}^{(3)} \ll \boldsymbol{L}^{(1)}\), unit permeability \(\boldsymbol{\mu}=\boldsymbol{I}\), and zero material magnetostriction \(\boldsymbol{M}^{mat}=\boldsymbol{0}\) to represent the magneto-mechanical properties of free space. These unit cells are denoted ``TetPv''.

A summary of the unit cells studied in this work and the abbreviations used to describe them is shown in Table~\ref{tbl:methods-abbrev}. Unless otherwise noted, all simulations were performed using the material properties shown in Table~\ref{tbl:mater-props}, which are chosen to be approximately representative of poly(dimethylsiloxane) (PDMS) elastomer matrix and carbonyl iron particles, respectively. All constituent materials were assumed to be isotropic and have no ``inherent'' material magnetostriction, i.e. the magneto-mechanical coupling tensor of each material phase is computed from (\ref{eq:M-decomposition}) assuming that \(M^{mat}_{ijkl} = 0\).

\begin{table}[H]
	\centering
	\caption{Notation describing the unit cell geometries studied in this work.} \label{tbl:methods-abbrev}
	\begin{tabu} to 0.9\textwidth {X[1.1] X[1.45] X[1] X[4.0]}
		\toprule
		Abbreviation          & Symmetry                   & Particle shape & Chain type                             \\ \midrule
		TetP             & Tetragonal (primitive)     & Spherical      & Straight                                   \\
		TetC             & Tetragonal (body-centered) & Spherical      & Straight                                   \\
		Hex              & Hexagonal                  & Spherical      & Straight                                   \\ \midrule
		TetPw            & Tetragonal (primitive)     & Spherical      & Zig-zag                                   \\
		TetPs-\(n\)      & Tetragonal (primitive)     & Spherical      & Spiral, \(n\) particles per turn                                   \\
		TetPv            & Tetragonal (primitive)     & Spherical      & Straight (void between particles)            \\ \midrule
		TetPe-\(\alpha\) & Tetragonal (primitive)     & Spheroidal     & Straight (particle minor radii ratio of \(\alpha\))   \\
		TetPc-\(i\)      & Tetragonal (primitive)     & Cylindrical    & Straight (particle axis parallel to \(x_i\)) \\ \bottomrule
	\end{tabu}
\end{table}

\begin{table}[H]
	\centering
	\caption{Material properties used in FEM simulations.} \label{tbl:mater-props}
	\begin{tabu} to 0.9\textwidth {X[2] X[1] X[1] X[1]}
		\toprule
		Property                       & Matrix       & Particle   & Void  \\
		\midrule
		Young's modulus, \(E\)         & 1.3 MPa      & 200 GPa    & 100 Pa \\
		Poisson's ratio, \(\nu\)       & 0.49         & 0.3        & 0.25 \\
		Relative permeability, \(\mu\) & 1            & 400        & 1 \\
		\bottomrule
	\end{tabu}
\end{table}

\section{Magneto-mechanical properties for straight-chain microstructures}
\label{sec:results.straight-chains}

The straight-chain microstructures possess only a small number of geometric parameters, allowing us to easily determine the effect of each parameter on the effective properties tensors, namely \(\tilde{\boldsymbol{L}}\), \(\tilde{\boldsymbol{\mu}}\), and \(\tilde{\boldsymbol{M}}\). In particular, we study the effects of individually varying the unit cell symmetry, volume fraction, gap size, and particle shape. Additionally, despite the simplicity of the straight-chain microstructures, we find that they reveal general trends in the effective properties of anisotropic MAEs that hold even for more complex unit cells.

\subsection{Dependence on unit cell symmetry}
\label{sec:results.straight-chains.symmetries}

\begin{figure}
    \centering
    \includegraphics{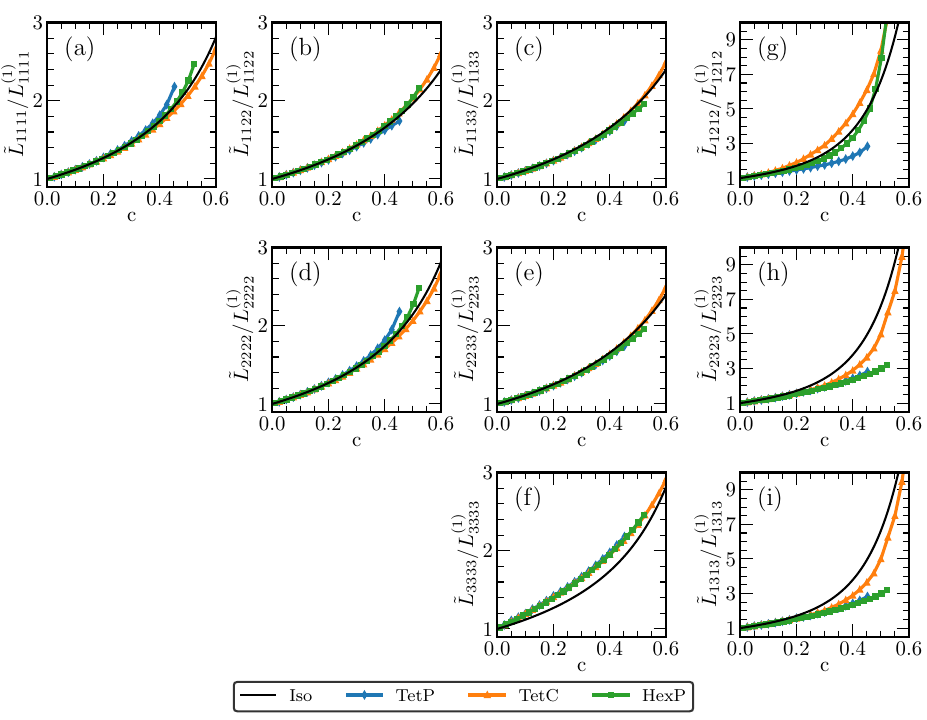}
    \caption{Nonzero components of the stiffness tensor for straight-chain geometries with fixed gap size \(d=0.05\), normalized by the corresponding components of the matrix stiffness tensor. Symmetric components are omitted.}
    \label{fig:straight-chain-L}
\end{figure}

To determine the effect of the unit cell symmetry, we compute the effective properties tensors for each symmetry with a fixed particle-to-particle gap (\(d=0.05\)) and volume fractions varying from \(c=0.025\) to the pseudo-percolation point \(c=c_{max}\). Figure~\ref{fig:straight-chain-L} shows the non-zero components of the stiffness tensor \(\tilde{\boldsymbol{L}}\) for the TetP, TetC, and Hex unit cells prior to the second (TI) homogenization step, with symmetric components excluded. In all plots, the isotropic model of \cite{Lefevre2014} is shown for comparison. For all volume fractions up to the onset of percolation, the longitudinal stiffness components (\(\tilde{L}_{1111}\), \(\tilde{L}_{2222}\), \(\tilde{L}_{3333}\), \(\tilde{L}_{1122}\), \(\tilde{L}_{2233}\), and \(\tilde{L}_{1133}\); Figure~\ref{fig:straight-chain-L}(a-f)) do not differ greatly between the TetP, TetC, and Hex unit cells, and with the exception of \(\tilde{L}_{3333}\) are also very close to the respective stiffness components for isotropic MAE. This indicates that the longitudinal stiffness components are not significantly affected by the artificial symmetry of the chosen unit cells, and thus can accurately represent the effective stiffness of a real anisotropic MAE. On the contrary, the shear stiffness components \(\tilde{L}_{1212}\), \(\tilde{L}_{2323}\), \(\tilde{L}_{1313}\) (Figure~\ref{fig:straight-chain-L}(g-i)) are highly dependent on the unit cell geometry under consideration. This is consistent with the suggestion of Hashin and Shtrikman \cite{Hashin1963} that the shear modulus is more dependent than the bulk modulus on the precise details of the composite microstructure. The dependence on the unit cell symmetry is especially strong for the \(\tilde{L}_{1212}\) component, which is expected since the unit cell symmetry primarily affects the arrangement of the particles in the \(xy\)-plane. We note that the shear stiffness components of the TetC geometry are generally greater than the corresponding components for the TetP and Hex geometries. This is due to the staggered arrangement of particles in adjacent chains in the TetC geometry, which creates an interlocking structure between chains that resists shear. Conversely, the TetP and HexP geometries have planes parallel to the \(z\)-axis which are uninterrupted by any particles, making them susceptible to shear in these planes. The \(\tilde{L}_{2323}\) and \(\tilde{L}_{1313}\) components are lower for all anisotropic geometries than for the isotropic solution. This is likely due to the fractal-like arrangement of particles in the differential coated sphere geometry for which the isotropic effective properties were obtained, which by their interlocking nature would tend to resist shear. Finally, we note that the effective stiffness in the direction of the particle chains (\(\tilde{L}_{3333}\)) is higher for the anisotropic geometries than for the isotropic solution. This result is intuitive, as the nearly-rigid particles create a stress concentration in the small gap between particles in a chain, inducing an increase in \(\tilde{L}_{3333}\) relative to the isotropic solution. In general, this is a drawback of anisotropic MAEs, as the increased stiffness relative to an isotropic MAE reduces the macroscopic tendency of the anisotropic MAE to be deformed by a magnetic field.

\begin{figure}
    \centering
    \includegraphics{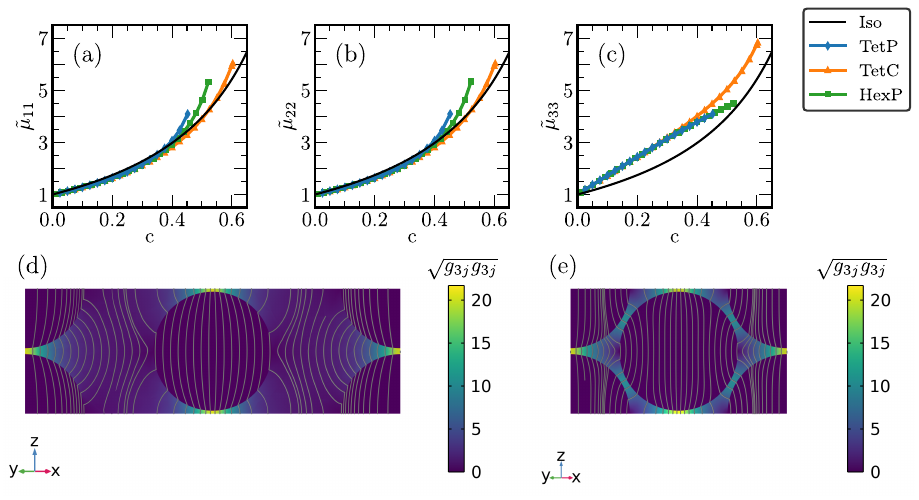}
    \caption{(a-c) Nonzero components of the effective relative permeability tensor for straight-chain geometries with fixed gap size \(d=0.05\). (d-e) Distribution of the magnetic field concentration tensor for the TetC geometry with (d) \(c=0.2\) and (e) \(c=0.6\). Streamlines indicate the local direction of the magnetic field and colors indicate the magnitude when the macroscopic field is in the \(z\)-direction.}
    \label{fig:straight-chain-mu}
\end{figure}

Figure~\ref{fig:straight-chain-mu} shows the non-zero components of the effective permeability tensor for the TetP, TetC, and Hex unit cells, again with \(d=0.05\) and \(0.025 < c < c_{max}\). Like the in-plane effective stiffness components, the in-plane effective permeability components (\(\tilde{\mu}_{11}\) and \(\tilde{\mu}_{22}\); Figure~\ref{fig:straight-chain-mu}(a-b)) are nearly equal between the three unit cells studied and are similar to the isotropic solution, for volume fractions up to percolation. Furthermore, \(\tilde{\mu}_{11} = \tilde{\mu}_{22}\) for each unit cell, implying that the effective permeability tensor has the TI symmetry expected for real anisotropic MAEs. The effective permeability in the direction of the particle chains (\(\tilde{\mu}_{33}\), Figure~\ref{fig:straight-chain-mu}(c)) is higher for the anisotropic MAEs than for the isotropic MAE, and \(\tilde{\mu}_{33}\) is consistent between the TetP, TetC, and Hex unit cells for volume fractions up to \(c=0.3\). Above \(c=0.3\), the TetC geometry exhibits a nonlinear increase in \(\tilde{\mu}_{33}\) which is not observed in the TetP and Hex geometries, due to an additional permeable path which arises at high volume fractions in the TetC unit cell from the overlapping of particles at the unit cell vertices and the particle at the body center. This allows some of the magnetic flux to jump back and forth between the body-centered particle and the vertex particles, as shown in the cross-section view in Figure~\ref{fig:straight-chain-mu}(e) where every magnetic flux line passing through the unit cell intersects at least one particle. In contrast, at low volume fractions in the TetC unit cell (Figure~\ref{fig:straight-chain-mu}(d)), and at all volume fractions in the TetP and Hex unit cells, there are magnetic flux lines that pass through the unit cell without intersecting a particle. This reveals that overlapping/staggered particle microstructures cause higher effective permeabilities. In nearly all cases, the permeability in the chain direction \(\tilde{\mu}_{33}\) is higher than that in the in-plane direction, with the exception of Hex unit cells with volume fractions near pseudo-percolation. In the Hex unit cells, \(\tilde{\mu}_{11}\) overtakes \(\tilde{\mu}_{33}\) at around \(c=0.49\), due to the Hex unit cell having closer packing in the \(xy\)-plane than in the chain (\(z\)-)~direction at high volume fractions.

\begin{figure}
	\centering
	\includegraphics{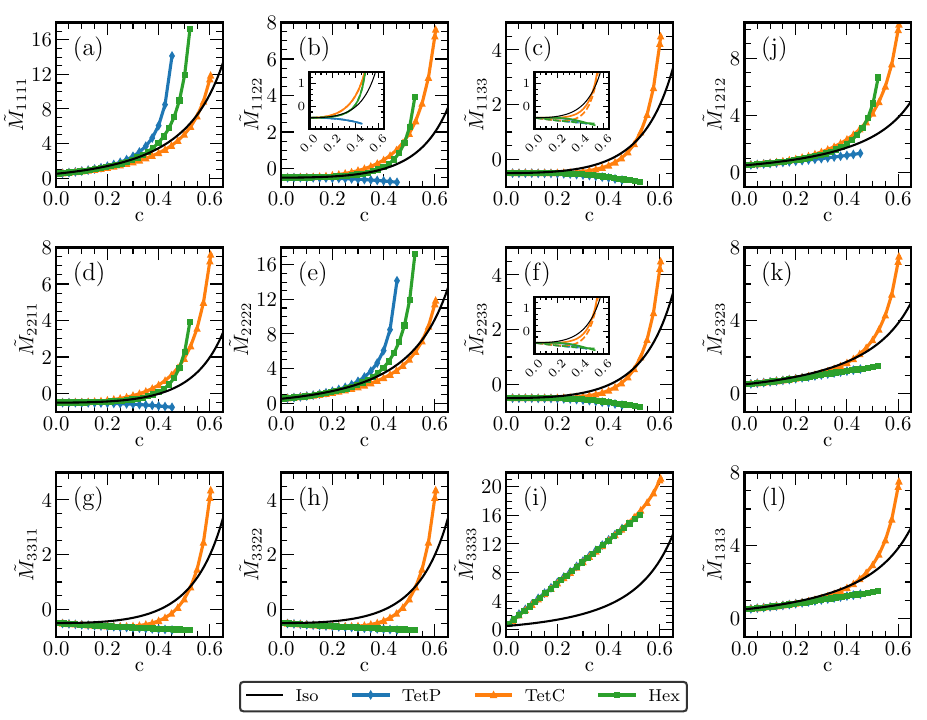}
	\caption{Nonzero components of the magnetostrictive coupling tensor for straight-chain geometries with fixed gap size \(d = 0.05\). Insets in (b, c, f) compare the off-diagonal components \(\tilde{M}_{1122}\), \(\tilde{M}_{1133}\), and \(\tilde{M}_{2233}\) (solid lines) with the corresponding transposed components, \(\tilde{M}_{2211}\), \(\tilde{M}_{3311}\), and \(\tilde{M}_{3322}\), respectively (dashed lines).}
	\label{fig:straight-chain-M}
\end{figure}

Figure~\ref{fig:straight-chain-M} shows the non-zero components of the magnetostrictive coupling tensor \(\tilde{\boldsymbol{M}}\) for the TetP, TetC, and Hex unit cells. It is immediately apparent that \(\tilde{\boldsymbol{M}}\) is more sensitive to the unit cell geometry than \(\tilde{\boldsymbol{L}}\) or \(\tilde{\boldsymbol{\mu}}\), as the components of \(\tilde{\boldsymbol{M}}\) differ more between different unit cells than do the components of \(\tilde{\boldsymbol{L}}\) and \(\tilde{\boldsymbol{\mu}}\). This is particularly evident from a comparison of the longitudinal magneto-mechanical coupling components (Figure~\ref{fig:straight-chain-M}(a-h)) with the longitudinal effective stiffness components (Figure~\ref{fig:straight-chain-L}(a-e)) and in-plane effective permeability components (Figure~\ref{fig:straight-chain-mu}(a-b)). Whereas the effective stiffness and permeability components are consistent between different unit cell symmetries up to around \(c=0.35\), the effective magnetostrictive coupling components (Figure~\ref{fig:straight-chain-M}a-h) are only consistent between unit cell symmetries up to a lower volume fraction of around \(c=0.2\). This is likely due to the quadratic dependence of \(\tilde{M}_{ijkl}\) upon the magnetic field concentration tensor: as the volume fraction is increased (decreasing the chain-to-chain spacing), the components \(G_{ijkl}\) and \(g_{ij}\) of the concentration tensors increase between the chains in a manner that is different for each unit cell due to the different arrangements of particle chains in the \(xy\)-plane. The difference between unit cells is magnified through the nonlinear dependence of \(\tilde{\boldsymbol{M}}\) upon \(\boldsymbol{G}\) and \(\boldsymbol{g}\), making \(\tilde{\boldsymbol{M}}\) more strongly dependent on unit cell symmetry than either \(\tilde{\boldsymbol{L}}\) or \(\tilde{\boldsymbol{\mu}}\), which depend only linearly on the concentration tensors. Since it is unclear which of the three unit cells---TetP, TetC, or Hex---would best represent a real MAE, caution should be exercised when using these results for volume fractions greater than \(c=0.2\), where the results begin to diverge between unit cells. Interestingly, the coupling coefficient in the particle chain direction (\(\tilde{M}_{3333}\), Figure~\ref{fig:straight-chain-M}(i)), is an exception to this rule and exhibits almost no dependence on the unit cell symmetry. This suggests that the results for \(\tilde{M}_{3333}\) could be valid for much larger volume fractions, up to around \(c=0.5\).

The off-diagonal components of \(\tilde{\boldsymbol{M}}\) (Figure~\ref{fig:straight-chain-M}(b-d, f-h)), which determine the coupling between a magnetic field in a given direction and the pre-stress in an orthogonal direction, show that such coupling is strongly dependent on the arrangement of the particles within the unit cell. In particular, the coupling is much stronger when the particles have a staggered arrangement with respect to the two directions being considered. For example, the TetC particles are arranged such that the nearest neighbors to a particle in the \(x\)- and \(y\)-directions are offset in the \(z\)-direction by half the unit cell thickness. This gives the TetC geometry strong couplings between the \(x\)/\(z\) and \(y\)/\(z\)-directions, manifested as large positive values of \(\tilde{M}_{1133}\) and \(\tilde{M}_{3311}\) for large volume fractions \(c\) where the chain-to-chain spacing is small enough to enable significant interactions between chains. In contrast, in the TetP and Hex geometries the nearest neighbors in the \(x\) and \(y\) directions are located at the same \(z\)-coordinate, and because of this non-staggered arrangement have weak \(x\)/\(z\) and \(y\)/\(z\) coupling. This is manifested as \(\tilde{M}_{1133}, \tilde{M}_{3311} \approx -0.5\) for all \(c\), which is nearly the same as the non-magnetostrictive elastomer matrix. Similarly, both the TetC and Hex geometries have a staggered arrangement in the \(xy\)-plane while the TetP geometry does not, resulting in TetC and Hex exhibiting larger values of \(\tilde{M}_{1122}\) and \(\tilde{M}_{2211}\) than TetP. A similar dependence on the staggering of particles in the unit cell is exhibited for the ``shear'' components \(\tilde{M}_{1212}\), \(\tilde{M}_{1313}\), and \(\tilde{M}_{2323}\) (Figure~\ref{fig:straight-chain-M}(j-l)), which determine the shear pre-stresses in response to macroscopic magnetic fields oblique to the coordinate axes. The effective magneto-mechanical coupling coefficient is much larger for unit cells with staggered particle arrangements, namely the TetC unit cell and the in-plane shear coupling (\(\tilde{M}_{1212}\)) of the Hex unit cell. It is likely that the TetC symmetry best captures the behavior of real anisotropic MAE, as electron micrographs typically reveal a staggered arrangement of particles within a chain in these materials, e.g. \cite{Boczkowska2012}.

The simulation results show that while the coupling tensor \(\tilde{\boldsymbol{M}}\) does not possess major symmetry, in practice it is nearly symmetric, as illustrated by the solid and dashed lines being nearly equal to each other in the insets of Figure~\ref{fig:straight-chain-M}(b, c, f). The \(\tilde{M}_{1122}\) and \(\tilde{M}_{2211}\) components are identical due to the symmetry of the TetP, TetC, and Hex unit cells (Figure~\ref{fig:straight-chain-M}(b) inset), while the \(\tilde{M}_{1133}\)/\(\tilde{M}_{3311}\) and \(\tilde{M}_{2233}\)/\(\tilde{M}_{3322}\) components exhibit a slight asymmetry (insets, Figure~\ref{fig:straight-chain-M}(c,f)). However, this asymmetry disappears during the TI homogenization step as we demonstrate in the following subsection, and thus may be considered to be the result of the artificial symmetries of the unit cell.

Most significantly, the simulations reveal a large enhancement in the magneto-mechanical coupling coefficient in the chain direction for anisotropic MAE relative to isotropic MAE, as shown in Figure~\ref{fig:straight-chain-M}(i). Whereas other components of the coupling tensor match the isotropic values for volume fractions as large as 0.3, \(\tilde{M}_{3333}\) diverges from the isotropic case starting at \(c=0\). The enhancement in \(\tilde{M}_{3333}\) appears to be much larger than the enhancement of \(\tilde{L}_{3333}\) (Figure~\ref{fig:straight-chain-L}(f)), which implies that anisotropic MAEs should exhibit higher magnetostriction than isotropic MAEs, despite being stiffer. Remarkably, \(\tilde{M}_{3333}\) appears to exhibit a linear relationship with volume fraction. The simplicity of this relationship suggests that it may be possible to find an analytical expression for \(\tilde{M}_{3333}\).

\subsubsection{Transversely isotropic effective properties}
\label{sec:results.straight-chains.symmetries.ti}

\begin{figure}
	\centering
	\includegraphics{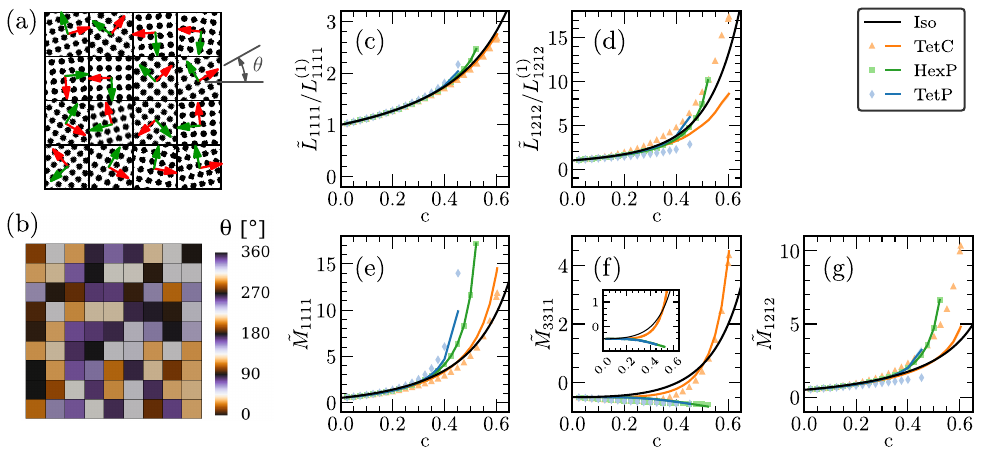}
	\caption{Transversely isotropic (TI) MAE effective properties for TetP, TetC, and Hex microstructures, computed from unit cells with grids of randomized \(xy\)-orientation. (a) schematic of \(4\times 4\) randomized unit cell and (b) implementation of a \(9 \times 9\) unit cell in COMSOL Multiphysics. (c-g) effective stiffness and effective magneto-mechanical coupling. Markers indicate effective properties prior to the TI homogenization step while lines indicate TI effective properties. Inset of (f) compares \(\tilde{M}_{3311}\) (solid lines) with \(\tilde{M}_{1133}\) (dashed lines); the curves are coincident for each unit cell.}
	\label{fig:straight-chain-TI}
\end{figure}

Figure~\ref{fig:straight-chain-TI} shows the independent components of the TI effective stiffness and magneto-mechanical coupling tensors obtained by applying the TI homogenization step to the results of the previous section. The \(\tilde{L}_{1133}\), \(\tilde{L}_{1313}\), \(\tilde{L}_{3333}\), \(\tilde{M}_{1133}\), \(\tilde{M}_{1313}\), and \(\tilde{M}_{3333}\) stiffness and coupling components are omitted because they are unchanged by the TI homogenization step. (Combining these components with those shown in Figure~\ref{fig:straight-chain-TI}, we have the five independent constants required to describe the major-symmetric TI stiffness tensor \(\tilde{\boldsymbol{L}}\), and the six independent constants required for the non-major-symmetric TI coupling tensor \(\tilde{\boldsymbol{M}}\).) The effective permeability is omitted from Figure~\ref{fig:straight-chain-TI} because the permeability tensor obtained from the first homogenization step (Figure~\ref{fig:straight-chain-mu}) already possesses TI symmetry and is thus unaffected by the TI homogenization step. In each plot of Figure~\ref{fig:straight-chain-TI}, the average value of 20 randomized geometries is shown with a solid line. The range of the results for each of the 20 geometries is less than the width of the line used to plot each effective property component. This indicates that the \(9 \times 9\) grid is sufficient to represent a fully random geometry. Figure~\ref{fig:straight-chain-TI}(c) shows that the in-plane longitudinal effective stiffness \(\tilde{L}_{1111}\) is not strongly affected by the TI homogenization step, while the in-plane shear stiffness \(\tilde{L}_{1212}\) is dramatically altered by the TI homogenization for the tetragonal unit cells. Figure~\ref{fig:straight-chain-TI}(d) reveals that the artificial symmetry of the TetP unit cell resulted in an underestimate of \(\tilde{L}_{1212}\), while the artificial symmetry of the TetC unit cell resulted in an overestimate. The \(\tilde{L}_{1212}\) component for Hex geometries is not affected, because hexagonal unit cells possess sufficient rotational symmetry to be transversely isotropic. The TI homogenization step brings both \(\tilde{L}_{1111}\) and \(\tilde{L}_{1212}\) closer to the isotropic properties, as expected. The components of the magneto-mechanical coupling tensor, shown in Figure~\ref{fig:straight-chain-TI}(e-g), are affected by the TI homogenization step in much the same manner as the effective stiffness components. The largest effect is noticed for the \(\tilde{M}_{1212}\) component (Figure~\ref{fig:straight-chain-TI}(g)), which is significantly increased by the TI homogenization step for TetP geometries, and significantly decreased for TetC geometries. Interestingly, the TI \(\tilde{M}_{1212}\) for the TetC geometry very closely approximates the isotropic geometry up to very high volume fractions of approximately \(c=0.55\).

While TI symmetry does not require \(\tilde{M}_{1133}=\tilde{M}_{3311}\), the TI homogenization step erases the slight asymmetry between these components that was shown in the inset of Figure~\ref{fig:straight-chain-M}(c). The TI homogenization step does not affect \(\tilde{M}_{1133}\), but increases \(\tilde{M}_{3311}\) such that it equals \(\tilde{M}_{1133}\) for all geometries, with the largest increase occurring for \(0.2<c<0.4\). Recalling that \(\tilde{M}_{1133}\) (\(\tilde{M}_{3311}\)) describes the 11- (33-) component of the pre-stress in the MAE when a magnetic field in the \(z\)- (\(x\)-) direction is applied, this implies that the pre-stress in the 11-direction under a \(z\)-directed magnetic field is equal to the pre-stress in the 33-direction under an \(x\)-directed magnetic field. This symmetry may be due to the high symmetry of the spherical particles used in these simulations, and may not exist for other particle shapes. Furthermore, while the magnetic pre-stresses in the two directions are equal under equal applied magnetic fields, it is expected that the magnetostrictions in the two directions will not be equal. In particular, the 11-strain under a \(z\)-directed magnetic field will be higher than the 33-strain under an \(x\)-directed magnetic field of equal strength, due to the lower stiffness \(\tilde{L}_{1111}\) of the MAE in the in-plane direction than in the chain direction (\(\tilde{L}_{3333}\)).

\subsection{Dependence on gap size}
\label{sec:results.straight-chain.gap-size}

\begin{figure}[H]
	\centering
	\includegraphics{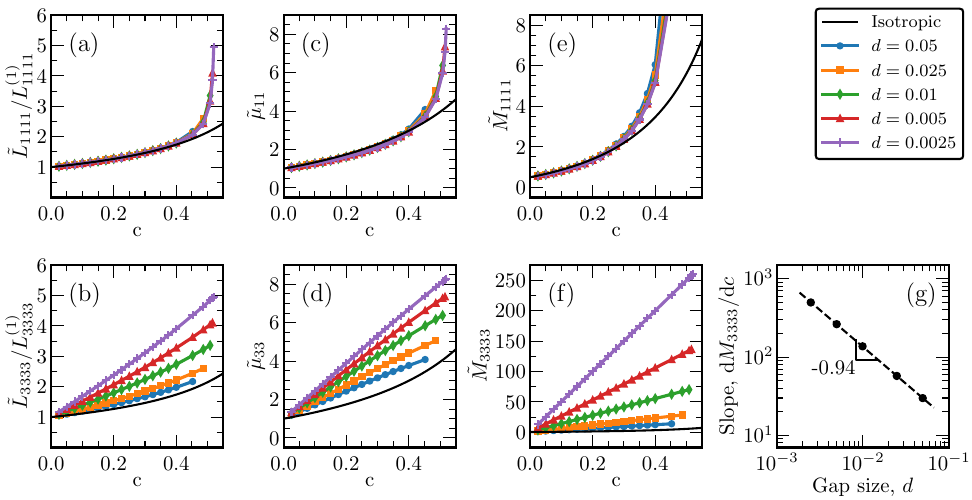}
	\caption{Effective properties of anisotropic MAEs with TetP microstructures for varying particle-to-particle gap: (a, b) effective stiffness, (c, d) effective permeability, (e, f) effective magneto-mechanical coupling. (a, c, e) components perpendicular to the particle chains; (b, d, f) components parallel to the particle chains. (g) variation of \(\tilde{M}_{3333}\) with gap size, \(d\).}
	\label{fig:eff-props-gap-size}
\end{figure}

The results presented in the previous section represent the effective properties of anisotropic MAEs with a fixed gap between adjacent particles in a chain. In this section, we examine the effect of gap size \(d\) on the effective properties. Since \(\tilde{L}_{3333}\), \(\tilde{\mu}_{33}\), and \(\tilde{M}_{3333}\) are the components most strongly enhanced by the anisotropy of the MAE, we focus on these components and provide selected in-plane components for comparison, presenting results only for the TetP geometry for clearness of exposition. As shown in Figure~\ref{fig:eff-props-gap-size}(a,c,e), the in-plane effective stiffness, permeability, and coupling coefficient have almost no dependence on the gap size, with slight differences observed only near percolation. The difference is due to the slight change in the aspect ratio of the unit cell induced by adjusting the gap size: decreasing the gap size decreases the length of the unit cell in the \(z\)-direction; thus, to maintain a fixed volume fraction \(c\), the \(x\)- and \(y\)- dimensions of the unit cell must increase slightly. This increased spacing between particle chains in the \(xy\)-plane causes a slight reduction in the in-plane components of the effective properties, but this effect may be regarded as an artifact of the construction of the unit cell geometry, and is unlikely to exist in real MAEs in which the chain-to-chain spacing is random.

In contrast to the in-plane properties, the effective properties in the particle chain direction (Figure~\ref{fig:eff-props-gap-size}(b, d, f)) are strongly affected by the gap size. As the gap size is decreased from 5\% to 0.25\% of the particle diameter (\(d=0.05\) to \(d=0.0025\)), the effective stiffness \(\tilde{L}_{3333}\) and permeability \(\tilde{\mu}_{33}\) approximately double. This is intuitive, as reducing the length of the low-stiffness (low-permeability) gap between particles increases the overall stiffness (permeability) of the composite. The most striking effect, however, is observed for the effective magneto-mechanical coupling coefficient \(\tilde{M}_{3333}\) (Figure~\ref{fig:eff-props-gap-size}(f)), which increases by more than an order of magnitude as \(d\) is varied over the same range. To quantify the growth in \(\tilde{M}_{3333}\), we calculate the slopes \(\mathrm{d}\tilde{M}_{3333}/\mathrm{d}c\) of the lines in Figure~\ref{fig:eff-props-gap-size}(f) using least-squares fits. Figure~\ref{fig:eff-props-gap-size}(g) shows the variation of the slope \(\mathrm{d}\tilde{M}_{3333}/\mathrm{d}c\) as a function of the gap parameter \(d\) for TetP geometries. The slopes appear to exhibit a power-law dependence on \(d\) with an exponent of -0.94. The same scaling law is also observed for both the TetC and Hex geometries. As the lines in Figure~\ref{fig:eff-props-gap-size} all intercept the \(\tilde{M}_{3333}\)-axis at 0.5, we obtain an empirical formula for \(\tilde{M}_{3333}\) as a function of volume fraction \(c\) and gap size \(d\) for microstructures comprising straight chains of spherical particles:
\begin{align}
	\tilde{M}_{3333} \approx 0.5 + \frac{\beta c}{d^{0.94}} \, , \label{eq:results-M3333-empirical}
\end{align}
where \(\beta\) is a fitting parameter that is equal to 1.79 for the material properties used in these simulations. Figure~\ref{fig:eff-props-gap-size}(f) clearly illustrates the advantage of anisotropic MAE over isotropic MAE, as the arrangement of particles into chains causes an enormous enhancement in \(\tilde{M}_{3333}\) while inducing only a modest increase in \(\tilde{L}_{3333}\), thereby increasing the magnetostriction of the MAE. This enhancement is achieved because the magnetic field applied during curing effectively induces percolation (contact between neighboring particles) in the \(z\)-direction regardless of volume fraction, and percolation is associated with dramatic changes in the effective properties of a composite. Thus anisotropic MAE can exhibit large values of \(\tilde{M}_{3333}\) even for small volume fractions, while in the isotropic MAE, percolation (and thus, large coupling coefficients) can occur only at large volume fractions.

Clearly, the expression (\ref{eq:results-M3333-empirical}) can only be valid for a limited range of gap sizes \(d\), since it implies that \(\tilde{M}_{3333}\rightarrow\infty\) as \(d \rightarrow 0\), i.e. the magnetostriction of anisotropic MAE should be infinite when the gap between particles is zero. This nonphysical result makes an analytical expression for \(\tilde{M}_{3333}\) highly desirable to probe the behavior of anisotropic MAE at vanishingly small gap sizes.

\subsection{Dependence on particle shape}
\label{sec:results.straight-chain.particle-shape}

The simulations discussed up to this point consider only spherical particles. When particles of other shapes are considered, namely cylinders and spheroids, \(\tilde{M}_{3333}\) retains its linear relationship with \(c\). We therefore quantify the relationship between \(\tilde{M}_{3333}\), volume fraction \(c\), and gap size \(d\) by computing the slopes \(\mathrm{d}\tilde{M}_{3333}/\mathrm{d}c\) as in the previous subsection, for all shapes studied. For unit cells comprising tetragonal (primitive)  arrangements of cylindrical particles (TetPc-2 and TetPc-3, Figure~\ref{fig:unit-cells}(f2-f3)) and spheroidal particles (TetPe-\(\alpha\), \(\alpha \in \{0.2, 0.5, 2.0, 5.0\}\)), each geometry exhibits an approximate power-law dependence on \(d\) over the range \(0.0025 \leq d \leq 0.05\), with a different exponent \(m\) for each particle type that can be generalized into the following empirical relationship:
\begin{align}
	\tilde{M}_{3333} \approx 0.5 + \frac{\beta c}{d^m} \, . \label{eq:results-M3333-empirical-power-law}
\end{align}
The TetPc-3 geometry exhibits the strongest scaling versus \(d\) (\(m=1.51\)), followed by the TetPc-2 geometry (\(m=1.33\)) and the TetP geometry (\(m=0.94\)). This suggests that for straight particle chains, the dependence of \(\mathrm{d}\tilde{M}_{3333}/\mathrm{d}c\) on \(d\) is determined by the average gap thickness between two adjacent particles in a chain, defined as the inter-particle gap volume divided by the projected area of the particle. For a fixed minimum gap \(2Rd\) between adjacent particles in the chain, the average gap thickness is governed by the curvature of the particle surfaces in contact. Flatter surfaces yield smaller average gaps and higher exponents \(m\): the TetPc-3 geometry has a flat (zero curvature) contact surface giving it the smallest average gap thickness (\(2Rd\)) and the strongest scaling (\(m=1.51\)), while ``curvier'' surfaces yield larger gaps and smaller \(m\): the TetP geometry (spherical particles) has a contact surface with curvature in two coordinate directions (\(x\) and \(y\)) giving it the largest average gap thickness (\(2Rd + \frac{2}{3}R\)) and the weakest scaling (\(m=0.94\)). The TetPc-2 geometry has a contact surface with curvature in one coordinate direction (\(y\)) giving it an average gap (\(2Rd + (2-\frac{\pi}{2})R\)) and scaling law (\(m=1.33\)) intermediate to the TetPc-3 and TetP geometries. The spheroidal particles are a generalization of the spherical particles, since the average gap thickness \(2Rd + \frac{2}{3}R\) is independent of the particle aspect ratio. Indeed, the scaling laws for the spheroidal particles have exponents \(m\) close to those for the spherical particles, with the different spheroidal aspect ratios having a stronger effect on the proportionality constant \(\beta\) than on \(m\). Prolate (oblate) particles, i.e. \(\alpha<1\) (\(\alpha > 1\)), yield higher (lower) coupling coefficients than spherical particles (\(\alpha = 1\)), which is due to the prolate particles producing a higher concentration of the magnetic field near their contact point, due to their elongated shape. We report the parameters \(\beta\) and \(m\) for all straight-chain geometries in Table~\ref{tbl:straight-chain-fits}. The strong dependence of the scaling exponent \(m\) on the particle shape suggests that particle shape could prove to be a powerful tool for improving the magneto-mechanical coupling of MAEs. Most MAE preparations reported in the literature to this point have utilized spherical particles, which our results show limits the magneto-mechanical coupling coefficient compared to other particle shapes. Preparing anisotropic MAE from cylindrical or needle-like particles could greatly improve the magneto-mechanical coupling compared to current formulations.

\begin{table}
	\centering
	\caption{Empirically-determined fit coefficients for Equation~(\ref{eq:results-M3333-empirical-power-law}) for the effective magneto-mechanical coupling of straight-chain geometries.} \label{tbl:straight-chain-fits}
	\begin{tabu} to 0.4\textwidth {X[2.5] X[1c] X[1c]}
	    \toprule
	    Geometry         & \(m\) & \(\beta\) \\ \midrule
	    TetP             & 0.94  & 1.79      \\
	    TetC             & 0.94  & 1.79      \\
	    Hex              & 0.94  & 1.79      \\ \midrule
	    TetPc-2          & 1.33  & 1.34      \\
	    TetPc-3          & 1.51  & 2.71      \\ \midrule
	    TetPe-0.2        & 0.78  & 4.94      \\
	    TetPe-0.5        & 0.91  & 2.23      \\
	    TetPe-1.0 (TetP) & 0.94  & 1.79      \\
	    TetPe-2.0        & 0.93  & 1.77      \\
	    TetPe-5.0        & 0.88  & 2.06      \\ \bottomrule
	\end{tabu}
\end{table}

The power-law relationship (\ref{eq:results-M3333-empirical-power-law}) holds for MAE where the properties of the constituent materials are chosen to be representative of elastomers and iron particles, i.e. \(\tilde{\boldsymbol{L}} \sim O(1\,\mathrm{MPa})\) and \(\mu=1\) for elastomers and \(\tilde{\boldsymbol{L}} \sim O(100\,\mathrm{GPa})\) and \(100 < \mu < 10000\) for iron, respectively, and for practically-relevant gap sizes \(0.0025 < d < 0.05\). We consider that \(d=0.0025\) is an approximate lower limit for the relevance of the homogenization model, since typical particles used in MAE fabrication have diameters on the order of \(1-10 \mu m\) and the minimum gap size studied (\(d=0.0025\)) would correspond to a particle-to-particle gap of \(2.5-25\,\mathrm{nm}\). Since this length is of the same scale as the polymer network making up the elastomer matrix \cite{Rath2016}, the continuum assumption breaks down and different physics are required to describe the effective behavior of the MAE in this case.

\section{An analytical expression for \(\tilde{M}_{3333}\)}
\label{sec:thry-M3333}

To better understand the relationship between the coupling coefficient \(\tilde{M}_{3333}\) and the microstructural parameters, especially the gap size \(d\), we seek an analytical expression for \(\tilde{M}_{3333}\) in terms of \(c\) and \(d\). We consider the TetPc-3 unit cell shown in Figure~\ref{fig:unit-cells}(f3), which has cylindrical particles of radius \(R\) and length \(L\) with the axis of each particle parallel to the axis of the chain (the global \(z\)-axis). The particle occupies volume \(Y_{(2)}\) and is considered to be isotropic with stiffness \(\boldsymbol{L}^{(2)}\), permeability \(\mu^{(2)}\), and coupling \(\boldsymbol{M}^{(2)}\). We prescribe a small fixed gap of \(D = Ld\) between particles in the \(z\)-direction, with \(0 < d \ll 1\). The unit cell length in the \(z\)-direction, \(L_z\), is thus \(L_z = (1 + d)L\). Upon prescribing the volume fraction of particles, \(c\), we can express the unit cell dimensions perpendicular to the chain as \(L_{xy} = R \sqrt{\pi/(1+d)c}\). The remainder of the unit cell, \(Y_{(1)}=Y \setminus Y_{(2)}\), is occupied by the elastomeric matrix with isotropic stiffness \(\boldsymbol{L}^{(1)}\), permeability \(\mu^{(1)}\), and coupling \(\boldsymbol{M}^{(1)}\). Since ferromagnetic fillers used in MAE are generally much more permeable than the elastomeric matrix material, i.e. \(\mu^{(2)} \gg \mu^{(1)}\), the magnetic flux is ``channeled'' through the particle chains when the global direction of the magnetic field is along the chain direction, i.e. the magnetic flux density \(\boldsymbol{B}\) in the particles and in the interparticle gap is nearly constant and parallel to the \(z\)-axis. Thus, by neglecting leakage flux around the particle-particle interface, the \(g_{33}\) component of the magnetic field concentration tensor in the particles and the interparticle gap can be determined as:
\begin{align}
	g_{33} =& \begin{cases}
		\frac{1 + d}{\alpha_\mu + d} & \forall \boldsymbol{y} \in Y_{ip} \\
		\left( \frac{1 + d}{\alpha_\mu + d} \right) \alpha_\mu & \forall \boldsymbol{y} \in Y_{(2)}
	\end{cases}\,, \label{eq:thry-h33}
	\intertext{where \(\alpha_\mu = \mu^{(1)}/\mu^{(2)}\) is the matrix-particle permeability ratio, and the ``inter-particle gap'' \(Y_{ip} \subset Y_{(1)}\) denotes the cylinder of elastomer matrix having radius \(R\) and height \(Ld\) that is co-axial with the particle chain and lies between two consecutive particles in the chain. For the mechanical problem, if the effect of the lateral constraint imposed by the surrounding matrix material is neglected, the strain concentration tensor can be analogously determined as:}
	G_{3333} =& \begin{cases}
		\frac{1 + d}{\alpha_L + d} & \forall \boldsymbol{y} \in Y_{ip} \\
		\left( \frac{1 + d}{\alpha_L + d} \right) \alpha_L & \forall \boldsymbol{y} \in Y_{(2)}
	\end{cases}\,, \label{eq:thry-H3333}
	\intertext{where \(\alpha_L = L_{3333}^{(1)}/L_{3333}^{(2)}\) is the matrix-particle stiffness ratio. We note that in the limiting case where the particles are treated as rigid and infinitely permeable, i.e. \(\alpha_\mu = \alpha_L = 0\), the concentration tensors reduce to the simple forms}
	g_{33} = G_{3333} =& \begin{cases}
		1 + \frac{1}{d} & \forall \boldsymbol{y} \in Y_{ip} \\
		0 & \forall \boldsymbol{y} \in Y_{(2)}
	\end{cases}\,. \label{eq:thry-rigid-infperm}
\end{align}

Given the small thickness of the interparticle gap (\(d \ll 1\)), it is apparent from (\ref{eq:thry-h33}) and (\ref{eq:thry-H3333}), and especially from (\ref{eq:thry-rigid-infperm}), that \(g_{33}, \, G_{3333} \gg 1\) in the interparticle gap and \(g_{33} , \, G_{3333} \approx 0\) within the particles. The FEM simulations further show that \(g_{33}, G_{3333}\sim O(1)\) in the remaining unit cell volume \(Y_{(1)}\setminus Y_{ip}\), and that \(g_{p3},G_{mn33}\approx 0\) throughout most of the unit cell for all \(m, n, p \neq 3\). Therefore, the effective magneto-mechanical coupling coefficient \(\tilde{M}_{3333}\), which is given by the following integral over the entire unit cell:
\begin{align}
	\tilde{M}_{3333} =& \frac{1}{|Y|} \int_{Y} {
		M_{lmpq}(\boldsymbol{y}) G_{lm33}(\boldsymbol{y}) g_{p3}(\boldsymbol{y}) g_{q3}(\boldsymbol{y}) \, \mathrm{d}Y
	}\, , \label{eq:thry-M33-post-proc}
\end{align}
is dominated by the integral over the inter-particle gap. Making use of Equations~(\ref{eq:thry-h33}) and~(\ref{eq:thry-H3333}) and the approximate values for all other \(g_{ij}, G_{ijkl}\) observed in the FEM simulations, Equation~(\ref{eq:thry-M33-post-proc}) can be approximated as
\begin{align}
	\tilde{M}_{3333} =& \frac{1}{|Y|} \int_{Y_{(1)} \setminus Y_{ip}} {
		M^{(1)}_{3333} \, \mathrm{d}Y
	} + \frac{1}{|Y|} \int_{Y_{ip}} {
		M^{(1)}_{3333} \left(\frac{1 + d}{\alpha_L + d}\right) \left(\frac{1 + d}{\alpha_\mu + d}\right)^2 \, \mathrm{d}Y
	} \nonumber \\
	&+ \frac{1}{|Y|} \int_{Y_{(2)}} {
		M^{(2)}_{3333} \left(\alpha_L \frac{1 + d}{\alpha_L + d}\right) \left(\alpha_\mu \frac{1 + d}{\alpha_\mu + d}\right)^2 \, \mathrm{d}Y
	} \, , \label{eq:thry-M33-post-proc-approx}
	\intertext{which, noting that the integrands are constant and considering the geometric identities \(|Y_{(1)}|=(1-c)|Y|\), \(|Y_{(2)}|=c|Y|\), and \(|Y_{ip}|=cd|Y|\), reduces to:}
	\tilde{M}_{3333} =& M_{3333}^{(1)} + \left(\left(\frac{1 + d}{\alpha_L + d}\right)\left(\frac{1 + d}{\alpha_\mu + d}\right)^2 \left(M^{(1)}_{3333} d + M_{3333}^{(2)} \alpha_L \alpha_\mu^2\right) - (1 + d) M_{3333}^{(1)}\right)c \,. \label{eq:thry-TetP-cylZ-M33}
\end{align}
We remark that the effective magneto-mechanical coupling (\ref{eq:thry-TetP-cylZ-M33}) has a linear relationship with the volume fraction, as expected from the simulation results, and that for \(c=0\) the effective magneto-mechanical coupling of the matrix material \(M_{3333}^{(1)}\) is recovered. We note several specializations of the expression (\ref{eq:thry-TetP-cylZ-M33}). First, for the limiting case of rigid, infinitely permeable particles (\(\alpha_L=\alpha_\mu=0\)), Equation~(\ref{eq:thry-TetP-cylZ-M33}) reduces to
\begin{align}
	\tilde{M}_{3333} =& M_{3333}^{(1)} + (1+d)\left(
		\left( \frac{1 + d}{d} \right)^2 - 1
	\right) M^{(1)}_{3333} c \,. \label{eq:thry-TetP-cylZ-M33-rigInf} \\
	\intertext{which, for \(d \ll 1\), is approximately equal to}
	\tilde{M}_{3333} \approx&\, M^{(1)}_{3333} + \frac{M^{(1)}_{3333} c}{d^2} \, , \label{eq:thry-TetP-cylZ-M33-rigInfApprox}
\end{align}
which is the same as (\ref{eq:results-M3333-empirical-power-law}) with \(\beta=\tilde{M}^{(1)}_{3333}\) and \(m=2\). Equation~\ref{eq:thry-TetP-cylZ-M33-rigInfApprox} suggests that decreasing the gap size \(d\) causes \(\tilde{M}_{3333}\) to increase without bound. However, this monotonic scaling is true only for composites with infinite contrast between the matrix and particle properties. When \(\alpha_L\) and \(\alpha_\mu\) are nonzero (i.e. the particle stiffness and permeability are finite), \(\tilde{M}_{3333}\) is bounded and exhibits a maximum for some \(d_{\max{(M)}} > 0\). The maximum value results from a balance between the magnitude of the concentration tensors and the volume of the inter-particle gap: as \(d\) is decreased from an initial value (e.g. \(d = 0.1\)), the second integrand in (\ref{eq:thry-M33-post-proc-approx}) scales approximately as \(1/d^3\), while the volume \(|Y_{ip}|\) over which the integral is computed scales as \(d\); thus the integral scales approximately as \(1/d^2\). (The scaling of the integrand is such because \(\alpha_\mu\) and \(\alpha_L\) are much smaller than 1 for typical MAE formulations, generally \(10^{-4} < \alpha_\mu < 10^{-2}\) and \(\alpha_L \sim O(10^{-5})\); thus \((1+d)/(\alpha+d)\approx 1/d\) when \(d \sim O(0.1)\).) However, when \(d\) becomes much smaller than the stiffness (permeability) ratio \(\alpha_L\) (\(\alpha_\mu\)), the strain (magnetic field) concentration tensor inside the inter-particle gap saturates at a constant value \(G_{3333}\approx 1/\alpha_L\) (\(g_{33}\approx 1/\alpha_\mu\)), while the inter-particle gap volume continues to decrease proportionally to the gap thickness \(d\). Thus, the effective coupling coefficient \(\tilde{M}_{3333}\) is approximately proportional to \(d\) for very small \(d\) below the saturation point. Thus, when \(d\) is below the saturation value, further decreases in the gap size will reduce rather than enhance \(\tilde{M}_{3333}\).

If we take the limit as \(d \rightarrow 0\) of~(\ref{eq:thry-TetP-cylZ-M33}), we obtain the effective coupling coefficient of a microstructure comprising infinite fibers in the \(z\)-direction:
\begin{align}
	\tilde{M}_{3333} \approx&\, (1 - c) M^{(1)}_{3333} + c M^{(2)}_{3333} \, , \label{eq:thry-fibers-M33}
\end{align}
since \(d=0\) corresponds to a chain with no gaps between the particles, i.e. a continuous fiber of ferromagnetic material. This implies that the effective coupling coefficient in the fiber direction is the volume average of the coupling coefficients of the matrix and fibers. Thus, for microstructures containing fiber inclusions, the effective magneto-mechanical coupling coefficient is strictly limited to lie between the effective coupling coefficients of the constituent materials. The same is not true of chains of particles. For example, in the TetPc-3 unit cell, \(\tilde{M}_{3333}\) exceeds \(\tilde{M}_{3333}^{(2)}\) for gap sizes as large as \(d=0.01\) (with \(c \approx 0.3\)) and for volume fractions as small as 0.025 (with \(d=0.0025\)). This demonstrates that anisotropic MAE comprising chains of particles hold a clear advantage in magnetostrictive capability over anisotropic MAE with fiber inclusions. Furthermore, this demonstrates that the low-permeability, low-stiffness gap between magnetic particles is the ``active'' volume of the MAE, that is, the magneto-mechanical interactions that govern the magnetostriction of MAEs take place primarily in the narrow gaps between particles.

\begin{figure}[]
	\centering
	\includegraphics{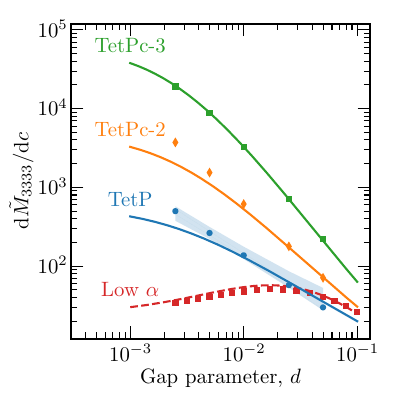}
	\caption{Dependence of effective magneto-mechanical coupling $\tilde{M}_{3333}$ on gap size \(d\) for tetragonal (primitive) geometries with different particle shapes. Markers indicate FEM simulations, lines indicate the theoretical model. Blue shaded area indicates the range of values obtained from FEM simulations of spheroidal particles with eccentricity ranging from 0.2 to 5.0. ``Low \(\alpha\)'' results are for TetPc-3 geometry with \(\alpha_L = \alpha_\mu = 0.04\), all other results are for \(\alpha_L = 8.3\times 10^{-5}\), \(\alpha_\mu = 1/400\).}
	\label{fig:slope-M-vs-d}
\end{figure}

For the general case (\ref{eq:thry-TetP-cylZ-M33}), we analyze the dependence of \(\tilde{M}_{3333}\) on \(d\) as in the previous section, by considering the derivative \(\mathrm{d}\tilde{M}_{3333}/\mathrm{d}c\), i.e.:
\begin{align}
		\frac{\mathrm{d}\tilde{M}_{3333}}{\mathrm{d}c} =& \left(\frac{1 + d}{\alpha_L + d}\right)\left(\frac{1 + d}{\alpha_\mu + d}\right)^2 \left(M^{(1)}_{3333} d + M_{3333}^{(2)} \alpha_L \alpha_\mu^2\right) - (1 + d) M_{3333}^{(1)} \,. \label{eq:thry-TetP-cylZ-M33-deriv}
\end{align}
In Figure~\ref{fig:slope-M-vs-d}, we plot (\ref{eq:thry-TetP-cylZ-M33-deriv}) for \(\alpha_L = 8.3\times 10^{-5}\) and \(\alpha_\mu = 1/400\), the values chosen for the simulations, along with the results from the simulations discussed in the preceding section. The expression (\ref{eq:thry-TetP-cylZ-M33-deriv}) is in excellent agreement with the simulation results for TetPc-3 geometries. The beginning of the saturation region is visible, with the curve of \(\mathrm{d}\tilde{M}_{3333}/\mathrm{d}c\) vs. \(d\) beginning to flatten out as \(d\) is decreased below around \(d=0.005\). To better illustrate the saturation behavior, we also plot (\ref{eq:thry-TetP-cylZ-M33-deriv}) for a TetPc-3 MAE with less contrast between the matrix and particle effective properties, \(\alpha_L = \alpha_\mu = 0.04\). When the material properties are chosen in such a way, the saturation of the effective magneto-mechanical coupling is clearly visible (Figure~\ref{fig:slope-M-vs-d}, ``Low \(\alpha\)'') and is confirmed by simulations. In this case, \(\tilde{M}_{3333}\) has a maximum for a gap size \(d_{\max(M)} \approx 0.018\) which is within the range of practically-relevant gap sizes \(0.0025 < d < 0.05\), indicating that to achieve the maximum coupling in such an MAE, it would be necessary to develop fabrication techniques to prevent the gap size from becoming too small. This is in contrast to MAE with iron particles, in which the gap size can in practice be decreased indefinitely without negatively impacting the coupling coefficient, since the maximum coupling coefficient occurs for \(d_{\max(M)}\) which are below the practically-relevant range of gap sizes.  The ``Low \(\alpha\)'' case could be representative of a ``hierarchical'' MAE where the particles are composed of MAE rather than pure iron, for example micron-sized MAE particles cured from droplets of a ferrofluid with nano-scale iron particles. In this hypothetical case, the constitutive properties of the particles would be typical of an isotropic MAE, i.e. the stiffness and permeability contrast \(\alpha_L\) and \(\alpha_\mu\) between the elastomer matrix and MAE particles would be \(O(0.1)-O(1)\).

The analytical results presented up to this point are valid for TetPc-3 geometries, but particles having different shapes and/or orientations may also be treated using a similar analysis, with appropriate corrections to account for the curvature of the particle surfaces in contact. We also considered the TetP and TetPc-2 unit cells, as shown in Figure~\ref{fig:unit-cells}(f1,f2), respectively. For TetPc-2 unit cells, the gap thickness depends on \(x\) according to the function \(t(x) = 2(1 + d)R - 2\sqrt{R^2 - x^2}\), where x is the distance from the chain axis in the x-direction. For TetP unit cells, the gap thickness is \(t(r)\), where \(r\) is the radial distance from the chain axis. The magnetic field and strain concentration tensors vary in space throughout the inter-particle gap and may be approximated as:
\begin{align}
	g_{33}(\xi) =& \frac{1 + \tau(\xi)}{\alpha_\mu + \tau(\xi)}\,, & G_{3333}(\xi) &= \frac{1 + \tau(\xi)}{\alpha_L + \tau(\xi)} & \forall \boldsymbol{y}& \in Y_{ip} \,, \label{eq:thry-H3333-sph}
\end{align}
where \(\tau(\xi) = t(\xi)/2R\)
is the normalized semi-thickness of the gap, with \(\xi=x\) for the TetPc-2 geometry and \(\xi=r\) for the TetP geometry. Replacing the constant values of the concentration tensors in (\ref{eq:thry-M33-post-proc}) with the spatially-varying equivalents (\ref{eq:thry-H3333-sph}) and making an appropriate substitution for \(\mathrm{d}Y\) in terms of \(x\) or \(r\), one can obtain analytical approximations for \(\tilde{M}_{3333}\) for TetPc-2 and TetP microstructures. However, the expressions are too lengthy to provide any meaningful insights and are not presented here. There is a poor fit between the theoretical model and the simulation results for TetP and TetPc-2 geometries, especially for small gap sizes \(d\). This is due to the assumption in the theoretical model that lines of magnetic flux remain parallel to the \(z\)-axis, even for particles with curved contact surfaces. The numerical simulations show that the curved face of the spherical particle bends the magnetic flux lines, causing \(g_{33}\) and \(G_{3333}\) to be larger near the center of the particle than what is predicted by (\ref{eq:thry-H3333-sph}). The theoretical model thus under-predicts the value of the coupling coefficient for the TetP and TetPc-2 geometries, as shown in Figure~\ref{fig:slope-M-vs-d}.

\section{Effect of other geometric parameters on magneto-mechanical properties} \label{sec:results.other-geoms}

While the results for straight-chain MAEs demonstrate that gap size and particle shape play key roles in determining the magneto-mechanical coupling coefficient, real anisotropic MAE microstructures contain many other geometric features not captured by the straight-chain models which may also affect the coupling coefficient. Here, we examine the effect on the effective properties of two features visible in microscope images of anisotropic MAEs: waviness of and voids in particle chains.

\subsection{Wavy chains} \label{sec:results.other-geoms.wavy}

\begin{figure}
    \centering
    \includegraphics{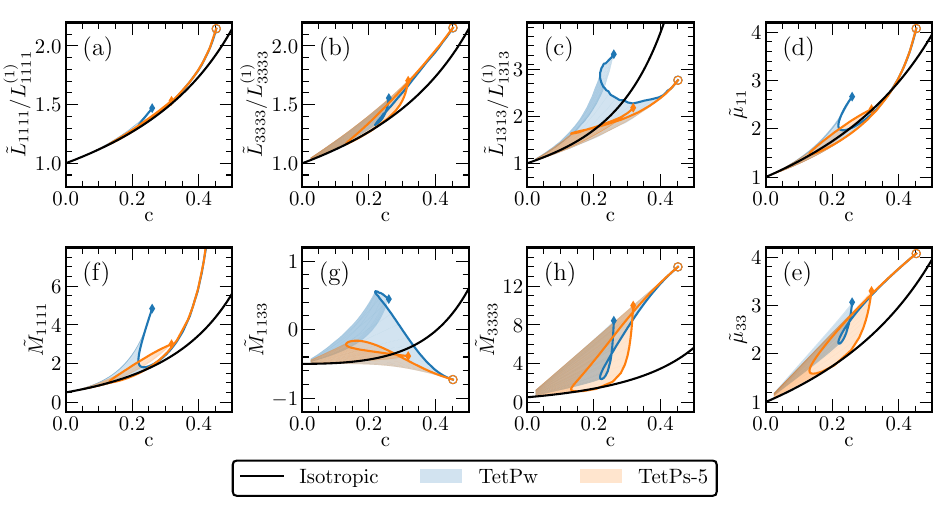}
    \caption{Selected components of the effective properties of wavy-chain MAEs with fixed gap size \(d=0.05\): (a-c) effective stiffness, (d-e) effective permeability, (f-h) effective magneto-mechanical coupling. Shaded areas indicate the accessible property space for all combinations of volume fraction and waviness parameter; solid lines indicate the effective properties at pseudo-percolation \(c_{max}\) with the circles indicating the zero-waviness (straight-chain) condition and diamonds indicating the maximum-waviness condition. Dark shading indicates self-overlapping regions of the effective property space.}
    \label{fig:wavy-chain}
\end{figure}

At a fixed gap size \(d\), the effective properties of wavy-chain anisotropic MAEs are parameterized by the volume fraction \(c\) and the parameter used to control the waviness. For the purposes of describing the wavy-chain results, we denote an arbitrary waviness parameter as \(\varphi\), where \(\varphi=\theta\) (the waviness angle) for TetPw unit cells and \(\varphi=a\) (the normalized spiral radius) for TetPs unit cells. \(\varphi = 0\) defines a straight chain, and increasing \(\varphi\) defines chains with increasing effective diameters, that is, the diameter of the cylinder parallel to the \(z\)-axis that circumscribes all particles in the chain increases. Since the unit cell length \(L_{xy}\) orthogonal to the particle chain must be greater than the effective chain diameter, the maximum volume fraction \(c_{max}\) which can be modeled is dependent on the waviness parameter and initially decreases with as \(\varphi\) increases from 0. Conversely, as the waviness parameter approaches its maximum value \(\varphi_{max}\), the chain ``collapses'' into a set of overlapping parallel chains with the packing density of the particles in the \(z\)-direction increasing rapidly, and \(c_{max}\) increases just before the waviness parameter reaches its maximum value. (For a visualization of the evolution of the chain geometry with increasing waviness, we refer the reader to the provided Supplementary Video 1.) For a fixed value \(\varphi = \varphi_0\) of the waviness parameter, the variation of an arbitrary effective property \(\tilde{X}\) with \(c\) describes a curve from \(\bigl(0, \tilde{X}(0)\bigr)\) to \(\bigl(c_{max}(\varphi_0), \tilde{X}(c_{max}(\varphi_0))\bigr)\) in \((c, \tilde{X})\)-space. As the waviness is smoothly varied from 0 to \(\varphi_{max}\), the curves \(\tilde{X}(c, \varphi)\) sweep out a region in (\(c\),\(\tilde{X}\))-space that defines the accessible effective property space for the wavy-chain MAEs, shown as transparently-shaded regions in Figure~\ref{fig:wavy-chain}. Material properties that are strongly affected by waviness undergo a large change as \(\varphi\) is varied from 0 to \(\varphi_{max}\) and the accessible property space spans a wide area, while a material property that is completely unaffected by the waviness would be represented by a set of coincident curves \(\tilde{X}(c, \varphi)\) which cover zero area. The set of points \(\tilde{X}(c_{max}(\varphi), \varphi)\), representing the effective property \(\tilde{X}\) evaluated at the pseudo-percolation volume fraction \(c_{max}\) for all \(0 \leq \varphi \leq \varphi_{max}\), defines a curve in \((c, \tilde{X})\)-space that is shown as a solid line in each plot of Figure~\ref{fig:wavy-chain}. These curves have hook-like shapes due to the non-monotonic dependence of \(c_{max}\) upon \(\varphi\) (\(c_{max}\) initially decreases as \(\varphi\) increases away from 0, then increases as the particle chain collapses near \(\varphi_{max}\)), with the shape of the hook determined by the dependence \(\tilde{X}(c_{max}(\varphi), \varphi)\) for a given property \(\tilde{X}\) and a particular type of waviness.

Of the three effective property tensors, the effective stiffness tensor (selected components shown in Figure~\ref{fig:wavy-chain}(a-c)) appears to be the least affected by waviness. The in-plane stiffness \(\tilde{L}_{1111}\) is almost completely unaffected by the waviness of the particle chain, as evidenced by the negligible shaded area in Figure~\ref{fig:wavy-chain}(a). Likewise, the off-diagonal stiffness components \(\tilde{L}_{1122}\), \(\tilde{L}_{1133}\), \(\tilde{L}_{2233}\) and the longitudinal stiffness \(\tilde{L}_{2222}\) (not shown) have almost no dependence on waviness. The stiffness \(\tilde{L}_{3333}\) in the chain direction is affected to a greater extent than \(\tilde{L}_{1111}\), appearing to be bounded from above by the straight-chain stiffness and from below by the isotropic stiffness. Still, the reduction in \(\tilde{L}_{3333}\) induced by waviness is relatively small, less than 15\% for all cases studied. The shear stiffness \(\tilde{L}_{1313}\) (Figure~\ref{fig:wavy-chain}(c)) is an exception; it has a wide accessible property space which spans the isotropic curve. This suggests that anisotropic MAEs with a random distribution of chain waviness would have an effective shear stiffness close to the isotropic shear stiffness, i.e. close to the center of the range shown in Figure~\ref{fig:wavy-chain}(c).

The effective permeability components, shown in Figure~\ref{fig:wavy-chain}(d-e), exhibit a modest dependence on waviness. The in-plane permeability (Figure~\ref{fig:wavy-chain}(d)) is mostly unaffected by the waviness parameter, with a fairly narrow accessible property space. The permeability in the chain direction, \(\tilde{\mu}_{33}\), however, is considerably reduced for intermediate to large waviness, with reductions of up to 50\% compared to straight chains. At maximum waviness \(\varphi_{max}\), \(\tilde{\mu}_{33}\) exceeds the straight-chain value, which is a result of the same particle-staggering mechanism that enhances \(\tilde{\mu}_{33}\) in TetC microstructures. As seen in e.g. Figure~\ref{fig:unit-cells}(e), a TetPs-\(n\) microstructure with \(a=a_{max}\) is equivalent to \(n\) parallel straight chains of particles, with staggering between chains, which results in a slight enhancement in \(\tilde{\mu}_{33}\) relative to the straight-chain case. These collapsed spiral chains bear some resemblance to realistic chains seen in microscope images of anisotropic MAEs, suggesting that the permeability of real MAEs may be similar to or slightly larger than the predicted permeability of straight-chain MAEs.

The effective magneto-mechanical coupling coefficients (Figure~\ref{fig:wavy-chain}(f-h)) depend strongly on the waviness, particularly the in-plane/out-of-plane coupling component \(\tilde{M}_{1133}\) and the out-of-plane component \(\tilde{M}_{3333}\). In the TetPw geometry, increasing waviness causes a change in sign of the \(\tilde{M}_{1133}\) coefficient; thus when a magnetic field is applied in the \(z\)-direction, pre-stresses in the 11-direction are tensile for highly wavy chains and compressive for straight and slightly wavy chains. This can be explained by the model of \cite{Biller2014} which shows that two particles magnetized by an external field will experience a force of attraction or repulsion that depends on the angle between the magnetic field and the vector connecting their centers. A straight chain of particles magnetized by a parallel magnetic field will form a chain of alternating north-south poles which are aligned side-by-side with the same poles in adjacent chains, generating a repulsive force between chains that causes the MAE to expand in the \(x\)-direction. This corresponds to the negative value of \(\tilde{M}_{1133}\) observed for straight chains, which gives a negative pre-stress that tends to induce tensile strains. Conversely, for zig-zag chains with \(\theta>14^\circ\), consecutive particles in the chain lie within the attraction region identified in \cite{Biller2014}, and the attractive force between the particles creates a tendency for the particles to align into a straight chain. This creates a net compressive strain in the \(x\)-direction, consistent with the positive value of \(\tilde{M}_{1133}\) (which corresponds to positive pre-stresses) for \(\theta>14^\circ\). This is qualitatively consistent with the particle movements noted by \cite{Kalina2016} for anisotropic MAEs.

The most significant impact of waviness is on the \(\tilde{M}_{3333}\) component. At zero waviness the straight-chain value is recovered, and for collapsed chains the strength of the coupling slightly exceeds the straight-chain value. However, for moderate waviness the enhancement achieved by arranging the particles in chains is almost completely eliminated. In the TetPw geometry with \(25^\circ < \theta < 58^\circ\), the excess of \(\tilde{M}_{3333}\) relative to isotropic MAEs is less than half that for straight chains. For TetPs-3 unit cells with \(a \approx 1\), \(\tilde{M}_{3333}\) is approximately equal to isotropic MAEs. This demonstrates that minimizing the waviness of particle chains is essential to achieve the full magnetostrictive potential of anisotropic MAEs.

\subsection{Void content} \label{sec:results.other-geoms.void}

\begin{figure}
	\centering
	\includegraphics{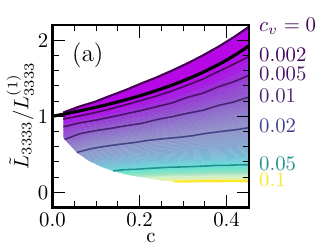}
	\caption{Effective stiffness in the chain direction for TetPv microstructures with \(d=0.05\). The colormap indicates the void volume fraction \(c_v\) with the scale indicated by the plotted contours. \(c_v = 0\) corresponds to the effective stiffness of TetP microstructures and the black line indicates the isotropic effective stiffness.}
	\label{fig:void}
\end{figure}

In previous sections, it was shown that increasing the magneto-mechanical coupling of the MAE came at the penalty of also increasing the effective stiffness, which limits the achievable macroscopic magnetostrictive response. Motivated by the appearance of voids in SEM images of MAE microstructures \cite{Chen2008}, caused by incomplete wetting of the particles by the matrix, we here study the effect of voids on the effective properties of anisotropic MAEs when the voids are located between the particles in a chain. Because the void has magnetic properties identical to the elastomer matrix, the effective permeability and effective magneto-mechanical coupling coefficient are unaffected by the introduction of voids into the MAE. Note that this behavior differs from DECs, in which the elastomer matrix typically has a dielectric constant greater than 1, with the result that voids tend to reduce the effective permittivity and electro-mechanical coupling of DECs. While the permeability and coupling of the MAE are unaffected by the void, the effective stiffness \(\tilde{L}_{3333}\) is significantly reduced by the introduction of voids between particles, as shown in Figure~\ref{fig:void}. The area spanned by the colormap in Figure~\ref{fig:void} indicates the variation of \(\tilde{L}_{3333}\) as the void volume fraction is increased from 0 to 10\%. As Figure~\ref{fig:void} shows, even a minute concentration of voids (0.2\%) is sufficient to reduce the anisotropic MAE stiffness to less than that of the isotropic MAE. For void content of 2\% or greater, the inclusion of the void completely negates the stiffness enhancement induced by the particles (\(\tilde{L}_{3333}/\tilde{L}^{(1)}_{3333} < 1\) for all \(c\)). For void content of 5\%, the effective stiffness is less than half that of the elastomer matrix, and between one-quarter and one-half that of the MAE with no voids. Since, as mentioned, this comes at no cost to the effective permeability and effective magneto-mechanical coupling, this indicates that anisotropic MAEs with strategic void placement could exhibit up to four times more magnetostriction than anisotropic MAEs with no voids. Even larger relative enhancements could be possible for MAEs with smaller gap sizes, since smaller gap sizes in anisotopic MAEs correspond to larger effective stiffnesses.

\section{Macroscopic magnetostrictive response of anisotropic MAEs} \label{sec:results-magnetostriction}

The preceding results demonstrate that \(\tilde{L}_{3333}\), \(\tilde{\mu}_{33}\), and \(\tilde{M}_{3333}\) are all enhanced by the chain-like particle microstructures. Increasing \(\tilde{M}_{3333}\) clearly increases the deformability of the MAE, while increasing \(\tilde{L}_{3333}\) decreases it. The effect of increasing \(\tilde{\mu}_{33}\) is not immediately clear. Therefore, to provide an overall picture of how the three effective property tensors interact to determine the magnetostriction in anisotropic MAEs, we compute the magnetically-induced strain induced in a uniform layer of anisotropic MAE. The MAE is taken to have finite thickness in the \(z\)-direction and infinite extent in the other directions, with the particle chains aligned through the thickness and a uniform magnetic flux \(B\) applied parallel to the chains. Under these loading conditions, the material undergoes a uniform strain \(\tilde{\varepsilon}\) in the \(z\)-direction due to the combined effect of the magnetic pre-stress \(\tilde{M}_{3333} (B/\tilde{\mu}_{33})^2 / \mu_0\) and the magnetic surface traction caused by the jump in the Maxwell stress at the MAE-air interface:
\begin{align}
	\tilde{\varepsilon} =& \frac{B^2}{\mu_0} \frac{1}{\tilde{L}_{3333}}\left(0.5 - \frac{\tilde{M}_{3333}}{{\tilde{\mu}_{33}}^2}\right) \, . \label{eq:thry-magnetostriction}
\end{align}
This differs from the analogous expression for DECs (see e.g., \cite{Lefevre2014}) due to the inclusion of the magnetic surface traction. This is a consequence of a subtle difference between DECs and MAEs: in DECs, it is standard practice to generate the electric field in the sample via parallel flexible electrodes bonded to the DEC surfaces, while in MAEs the magnetic field is usually generated by a rigid electromagnet. Thus in DECs the surface traction is zero while in MAEs it is nonzero. Equation~(\ref{eq:thry-magnetostriction}) shows that there is a competition between the \emph{tensile} magnetic traction on the top and bottom surfaces of the MAE and the magnetically-induced pre-stress in the MAE volume (which tends to induce \emph{compressive} strains), with the term in parentheses determining the direction of the deformation: elongation when \(\tilde{M}_{3333}/\tilde{\mu}_{33}^2 < 0.5\) and compression when \(\tilde{M}_{3333}/\tilde{\mu}_{33}^2 > 0.5\). This also implies that it could be possible to create an MAE which does not deform under the magnetic loading studied, if \(\tilde{M}_{3333}/\tilde{\mu}_{33}^2 = 0.5\).

\begin{figure}
    \centering
    \includegraphics{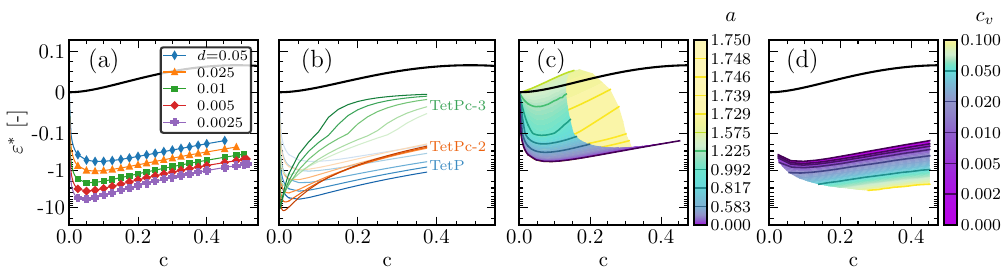}
    \caption{Magnetostriction \(\varepsilon^* = \mu_0 L_{3333}^{(1)} \tilde{\varepsilon} / B^2\) for (a) TetP unit cell with varying gap size \(d\), (b) different particle shapes, (c) TetPs-5 spiral geometry with varying spiral radius \(a\), and (d) TetPv geometry with \(d=0.05\) and varying void volume fraction \(c_v\). Isotropic MAE is shown in each plot with a solid black line. Note the split log scale of the ordinate in each subplot: strains \(|\varepsilon^*| \leq 0.1\) are displayed on a linear scale, strains \(|\varepsilon^*| > 0.1\) on a log scale.}
    \label{fig:macro-TetP-d}
\end{figure}

As shown in Figure~\ref{fig:macro-TetP-d}, isotropic MAEs tend to undergo tensile strain while anisotropic MAEs undergo compressive strain for most values of the parameters studied. This indicates that the MAE microstructure controls which mechanism drives magnetostriction: anisotropic MAEs are dominated by the magnetic pre-stress while isotropic MAEs are dominated by the magnetic surface traction. This behavior can be explained by considering the particles in the magnetized MAE as microscopic magnets. In anisotropic MAEs, the magnets are arranged in chains with their opposite poles aligned together, creating attractive forces along the chain that tends to compress the material. Due to the small size of the gap between the particles, this force is strong enough to overcome the tensile force on the MAE surfaces, and the MAE undergoes an overall compressive strain. In contrast, isotropic MAEs have a random arrangement of particles, so that some particles are attracted to each other while others are repelled from each other. Furthermore, the gap between particles is relatively larger in isotropic MAEs than in anisotropic MAEs at the same volume fraction, reducing the strength of the inter-particle forces in isotropic MAEs relative to anisotropic MAEs. The net result is that the inter-particle attractions are not strong enough to overcome the magnetic traction on the MAE surfaces, so the MAE elongates. The sign of the magnetostriction may therefore depend on whether the microstructure is close to percolation or not: the anisotropic MAEs studied in this work have particles nearly in contact along the chain direction and exhibit compressive strains, while the isotropic MAEs (\cite{Lefevre2014}) have a differential coated sphere microstructure that does not percolate until \(c=1.0\) and exhibit tensile strains. It is therefore of interest to study whether isotropic MAEs having a monodisperse particle size distribution (which percolate at volume fractions \(c<1.0\)) would exhibit compressive strains at some volume fraction near percolation.

The different driving mechanisms lead to qualitative differences in the magnetostrictive behavior of isotropic MAEs and anisotropic MAEs. As shown in Figure~\ref{fig:macro-TetP-d}(a), the maximum absolute value of strain is achieved at small volume fractions in the anisotropic MAEs, while it is achieved at a much higher volume fraction \(c\approx 0.45\) in the isotropic MAE. In the anisotropic MAEs, the maximum is determined primarily by a competition between \(\tilde{M}_{3333}\) and \(\tilde{\mu}_{33}\), through the ratio \(\tilde{M}_{3333}/\tilde{\mu}_{33}\) appearing in (\ref{eq:thry-magnetostriction}). For spherical particles (Figure~\ref{fig:macro-TetP-d}(a)), \(\tilde{M}_{3333}\) increases faster than \(\tilde{\mu}_{33}^2\) at small volume fractions, leading to an increase in the overall magnetostriction up to around \(c=0.1\). Above this point, the growth in \(\tilde{\mu}_{33}^2\) outstrips the growth in \(\tilde{M}_{3333}\); thus the magnetostriction in anisotropic MAEs is maximized at around \(c=0.1\). In the isotropic MAE, the maximum strain is determined primarily by the growth in stiffness: the ratio \(\tilde{M}_{3333}/\tilde{\mu}_{33}^2\) is monotonically decreasing, which would lead to a monotonic increase in the magnetostriction were it not for the nonlinear growth in  \(\tilde{L}_{3333}\) which causes the strain to be maximized at around \(c=0.45\).

As shown in Figure~\ref{fig:macro-TetP-d}, anisotropic MAEs could exhibit exponentially larger magnetostriction than isotropic MAEs, but this is strongly dependent on the microstructural parameters. Predictably, decreasing gap size increases the magnetostriction (Figure~\ref{fig:macro-TetP-d}(a)), with the absolute value of the strain increasing by more than an order of magnitude as \(d\) is decreased from 0.05 to 0.0025. This is in accordance with the observation in Section~\ref{sec:results.straight-chain.gap-size} that \(\tilde{M}_{3333}\) increases much faster than \(\tilde{L}_{3333}\) and \(\tilde{\mu}_{33}\) with decreasing gap size. For unit cells having spherical particles, the trend of increasing magnetostriction with decreasing gap size is independent of the volume fraction. However, for different particle shapes (Figure~\ref{fig:macro-TetP-d}(b)), the effect of gap size depends both on the particle shape and the volume fraction. For example, in TetPc-3 unit cells with low volume fractions, decreasing gap size is associated with higher magnetostriction, while the opposite relationship holds at higher volume fractions. The optimum particle shape (that which produces the largest magnetostriction) is also dependent on volume fraction and gap size: for volume fractions close to 0, TetPc-3 unit cells produce the highest magnetostriction, but are quickly overtaken by TetPc-2 unit cells as volume fraction is increased, while spherical particles yield the largest magnetostriction at high volume fractions. Thus, while cylindrical particles may yield orders-of-magnitude larger coupling coefficients than spherical particles, they also yield simultaneous increments of stiffness and permeability that negate the benefit of the higher coupling coefficient. Although not shown in Figure~\ref{fig:macro-TetP-d}, our simulations show that both prolate and oblate spheroidal particles exhibit similar but smaller magnetostrictions than spherical particles. It therefore appears that spheres are generally the most effective particle shape for a broad range of anisotropic MAE microstructures, a fortuitous result since most commercially available iron powders consist of spherical particles.

Chain waviness can completely negate the magnetostrictive advantage of anisotropic MAEs over isotropic MAEs, as shown in Figure~\ref{fig:macro-TetP-d}(c) for TetPs-5 (spiral) geometries with \(d=0.05\). The shaded area in this figure represents the range of strains which can be achieved over all allowable values of the spiral radius \(a\). The color of the shading indicates the value of \(a\) for which a particular strain is achieved, with several values of \(a\) marked by colored contour lines and labeled on the color bar. The shaded region is self-overlapping; that is, the region \(a \approx 1.75\) (shaded in yellow) extends to the origin (\(c=0\), \(\varepsilon_{33}=0\)) of the plot, but is not visible since it is plotted beneath the region of low \(a\) (shaded purple to green). In the obscured portion of the plot, the yellow contours follow trajectories towards the origin that are similar to the curves for \(a < 1\). The impact of chain waviness on the macroscopic response is significant. For example, \(a=0.583\) is a spiral with radius just greater than half the particle radius, which appears nearly straight. However, even for this nearly-straight chain, the strain is decreased by up to half compared to the straight chains. For chains with large waviness (e.g. \(1.225 < a < 1.729\)), the effective strain of the anisotropic MAE is on the same order of magnitude as the isotropic MAE, thereby eliminating the advantage of anisotropic MAEs. For spiral radii \(1.729 < a < 1.75\), the spiral collapses into a ring of five parallel particle chains, and a macroscopic response similar to that of the straight-chain geometries is recovered. This indicates that controlling the chain waviness is critical to maximizing the magnetostrictive response of anisotropic MAEs.

Conversely, voids in the particle chain can provide orders-of-magnitude enhancement in the MAE magnetostriction. As shown in Figure~\ref{fig:macro-TetP-d}(d), increasing void content \(c_v\) from 0 to 10\% increases the magnetostriction by greater than an order of magnitude relative to TetP geometries with no voids. The enhancement in magnetostriction is achieved because the voids reduce the MAE stiffness \(\tilde{L}_{3333}\), in some cases resulting in MAEs that are more compliant than the matrix for sufficiently large \(c_v\). The results shown in Figure~\ref{fig:macro-TetP-d}(d) are for relatively large gap sizes (\(d=0.05\)), and larger relative enhancements in the magnetostriction are possible for smaller gap sizes where the mechanical reinforcement due to the particles is larger. For example, TetP geometries with \(d=0.05\) have a maximum \(\tilde{L}_{3333}\) of approximately twice the matrix stiffness \(L^{(1)}_{3333}\) while TetP geometries with \(d=0.01\) can be up to 3.5 times stiffer than the matrix. Our simulations show that including voids in the chain at a concentration of \(c_v = 0.1\) effectively eliminates the reinforcing effect of the particles irrespective of gap size, reducing \(\tilde{L}_{3333}\) to the same value or smaller than the matrix stiffness \(L^{(1)}_{3333}\) without impacting \(\tilde{\mu}_{33}\) or \(\tilde{M}_{3333}\). Significantly larger magnetostriction can therefore be realized for gap size \(d=0.01\) than for \(d=0.05\), since these two unit cells have approximately the same effective stiffness but the smaller gap size has significantly larger coupling.

\section{Discussion}
\label{sec:discussion}

The constitutive model adopted in this work is strictly valid only in the case where the strains and magnetic fields throughout the unit cell are sufficiently small: \(O(\zeta)\) and \(O(\zeta^{1/2})\), respectively. While these conditions may prevail in isotropic MAE for fairly large macroscopic strains due to the relatively large gaps between particles, in anisotropic MAE these results may be severely limited in terms of the macroscopic strain for which they are applicable, due to the high strain concentration between the closely-spaced particles. For example, strains in the interparticle gap could be up to 400 times larger than the macroscopic strain for deformations along the chain axis when \(d=0.0025\), making these results strictly valid for macroscopic strains only on the order of \(10^{-5}\). It is therefore of interest to compute the effective properties of anisotropic MAE under larger strains incorporating nonlinearity, and study the effect of nonlinearity on the macroscopic magnetostriction. It could be expected that the nonlinearity would increase the effective stiffness of the MAE thereby reducing its overall magnetostriction, while it is also conceivable that compressing the MAE could increase the strength of magnetic forces between particles, increasing the overall magnetostriction.

Further, the proposed homogenization models ignore any interphasial layer in the elastomer surrounding the particles, which is known to exist and has been suggested to affect the effective properties of the composite \cite{Lefevre2017b}. Additionally, while some of the effects of chain waviness have been captured by our model, realistic particle chains in MAEs possess a greater degree of disorder than those studied in this work, for example with waviness over distances many times the particle diameter and interruptions in particle chains. Comparing our analytical and simulation results to experimental measurements of the macroscopic magnetostriction would inform model improvements and determine to what extent real MAE are affected by these other factors.

Achieving the large magnetostrictions demonstrated in this work would require control over microstructural parameters such as the gap size and the waviness that may present challenges to implement; however, the results in this paper motivate exploring fabrication techniques that could realize these specific microstructures. Control over the average gap between particles could possibly be obtained through the right combination of curing field strength, elastomer viscosity, and curing rate, or by using particles pre-coated with a thin elastomer layer. Inducing particles to align in straight chains rather than wavy ones might present an even greater fabrication challenge but could result in stronger coupling than what is possible with current MAEs. It has been shown that rotating magnetic fields produce layer-like arrangements of particles instead of the chains produced by static fields \cite{Liu2013a}, so it may be possible that other types of non-static magnetic fields could produce straighter chains than static fields.

The directionality of the strain shown in Figure~\ref{fig:macro-TetP-d} (positive for isotropic MAE, negative for anisotropic MAE) could have important ramifications for magnetostrictive actuators. A device in which the composition of the MAE was spatially varied (i.e. isotropic in some regions and anisotropic in others) could simultaneously exert tensile and compressive forces in response to a single magnetic field. Strategic patterning of isotropic and anisotropic regions in a thin MAE layer could be used, for example, in a haptic device capable of forming a textured surface with a desired pattern in response to an applied magnetic field. Design of such devices could be facilitated by the results presented in this work which serve as a ``toolbox'' of material properties that can be readily implemented in finite element software to analyze the performance of proposed device designs. The model could be employed in optimization routines with the microstructural parameters (e.g. volume fraction, chain direction) acting as design variables, where the pre-computed MAE effective properties would be used to solve the forward problem at each iteration of the optimization routine. The one-way coupled constitutive law adopted in this work would reduce the forward problem to two sequential linear problems which could be solved rapidly and reliably, in contrast to other fully nonlinear (possibly implicit) MAE constitutive models.

\section{Conclusions}

In conclusion, we have developed a computational homogenization framework which enables us to study the microstructural parameters that govern the magneto-mechanical coupling of anisotropic magneto-active elastomers, namely the effective stiffness and permeability tensors and the fourth-order tensor that couples magnetic fields to magnetically-driven deformations. Our simulation results show that anisotropic MAE potentially exhibit magnetostrictions orders of magnitude larger than what is possible with isotropic MAE. Guided by insights obtained from the full-field simulation results, we derived a fully explicit expression for the magneto-mechanical coupling coefficient in the particle chain direction in terms of the particle volume fraction, the properties of the matrix and particles, and the gap between particles in a chain. The analytical model, validated by the finite element simulations, shows that reducing the gap between particles in a chain significantly increases the overall magnetostriction by increasing the strength of the magnetic interactions between adjacent particles, while the presence of waviness in the particle chain can significantly reduce the magnetostriction. Our models further show that voids between particles in the chain can further increase the magnetostriction by reducing the overall MAE stiffness, and that spheres appear to be the best particle shape under most conditions because they strike the best balance between increased magneto-mechanical coupling and increased stiffness. Interestingly, our results show that while isotropic MAE are expected to elongate in a uniform magnetic field, anisotropic MAE compress, due to the closely-spaced particles in a chain experiencing much stronger attraction than the relatively widely-spaced particles in isotropic MAE. Our results provide a deeper fundamental understanding of the microstructural parameters which govern the magneto-mechanical coupling in anisotropic MAE, and could serve as a ``toolbox'' of material properties which can be used to design and analyze magnetostrictive devices.

\section*{Acknowledgments}

This work made use of the Illinois Campus Cluster, a computing resource that is operated by the Illinois Campus Cluster Program (ICCP) in conjunction with the National Center for Supercomputing Applications (NCSA) and which is supported by funds from the University of Illinois at Urbana-Champaign. C.D. Pierce acknowledges support received through the National Defense Science and Engineering Graduate Fellowship through the U.S. Department of Defense.

\section*{Funding}

This work was funded by the United States Air Force Office of Scientific Research under award number FA9550-20-1-0036. The funding agency did not participate in conducting the investigation, writing this manuscript, or in the decision to submit this manuscript for publication.

\section*{Author contributions}

\textbf{C. D. Pierce:} conceptualization, formal analysis, investigation, methodology, software, visualization, writing---original draft, writing---review \& editing. \textbf{K.H. Matlack:} conceptualization, funding acquisition, methodology, resources, supervision, writing---original draft, writing---review \& editing.

\printbibliography

\end{document}